\documentclass[twocolumn]{openjournal}
\usepackage{graphicx}
\usepackage{orcidlink}

\usepackage{kpfonts}
\begin{document}

\title{The Third Option: Color Phase Curves to Characterize the Atmospheres of Temperate Rocky Exoplanets}
\shorttitle{Color phase curves}
\shortauthors{Deming et al.}
\author{Drake Deming\orcidlink{0000-0001-5727-4094}}
\email{ldeming@umd.edu}
\affiliation{Department of Astronomy, University of Maryland,
   College Park, MD 20742, USA}
\affiliation{NASA's Nexus for Exoplanet System Science, Virtual Planetary Laboratory Team, Box 351580, University of Washington, Seattle, WA 98195, USA}  

\author{Andrew Lincowski\orcidlink{0000-0003-0429-9487}}
\affiliation{SETI Institute, Carl Sagan Center, 339 Bernardo Ave, Suite 200, Mountain View, CA 94043, USA}
\affiliation{NASA's Nexus for Exoplanet System Science, Virtual Planetary Laboratory Team, Box 351580, University of Washington, Seattle, WA 98195, USA}  

\author{Laura Kreidberg\orcidlink{0000-0003-0514-1147}}
\affiliation{Max-Planck-Institut für Astronomie, Königstuhl 17, D-69117 Heidelberg, Germany}

\author{Miles H. Currie\orcidlink{0000-0003-3429-4142}}
\affiliation{NASA Goddard Space Flight Center, Greenbelt, MD 20771, USA}

\author{Jean-Michel Desert\orcidlink{0000-0002-0875-8401}}
\affiliation{Leibniz Institute for Astrophysics, An der Sternwarte 16, 14482 Potsdam, Germany}

\author{Guangwei Fu\orcidlink{0000-0002-3263-2251}}
\affiliation{Department of Physics and Astronomy, Johns Hopkins University, Baltimore, MD, USA}

\author{Jacob Lustig-Yaeger\orcidlink{0000-0002-0746-1980}}
\affiliation{JHU Applied Physics Laboratory, 11100 Johns Hopkins Rd, Laurel, MD 20723, USA}

\author{Victoria S. Meadows\orcidlink{0000-0002-1386-1710}}
\affiliation{SETI Institute, Carl Sagan Center, 339 Bernardo Ave, Suite 200, Mountain View, CA 94043, USA}
\affiliation{Department of Astronomy and Astrobiology Program, University of Washington, Box 351580, Seattle, WA 98195, USA}
\affiliation{NASA's Nexus for Exoplanet System Science, Virtual Planetary Laboratory Team, Box 351580, University of Washington, Seattle, WA 98195, USA}

\author{Ignas Snellen\orcidlink{0000-0003-1624-3667}}
\affiliation{Leiden Observatory, Leiden University, Postbus 9513, 2300 RA, Leiden, The Netherlands}

\begin{abstract}
Detecting and characterizing the atmospheres of rocky exoplanets has proven to be challenging for JWST.  Transit spectroscopy of the TRAPPIST-1 planets has been impacted by the effects of spots and faculae on the host star.  Secondary eclipses have detected hot rocks, but evidence for atmospheres has been difficult to obtain.  However, there is a third option that we call color phase curves.  This method will apply to synchronously rotating non-transiting planets as well as transiting planets.  A color phase curve uses photometry at a long-IR wavelength where the planetary thermal emission is strong (e.g., 21 microns) divided by photometry at a shorter wavelength where the star dominates (e.g., 12 microns).  We avoid wavelengths having potentially strong molecular absorption (e.g., 15 microns) to minimize degeneracies in the color phase curve, and we aim to detect and characterize the planetary atmosphere via its longitudinal heat transfer.  The ratio of two wavelengths observed nearly simultaneously is designed to isolate thermal emission from the planet, discriminate against the star, and largely cancel instrumental systematic effects.  Moreover, we show that invoking mass-radius relations, and using self-consistent physical models, will permit the longitudinal heat transfer to be measured independent of the orbital inclination.  Radial velocity surveys are detecting many new exoplanets, including temperate rocky worlds with Earth-like masses. Most of those planets will not transit, but color phase curves have the potential to detect and characterize their atmospheres. 
\\
\\
\end{abstract}


\section{Introduction}\label{sec: intro}
Detecting and characterizing the atmospheres of rocky exoplanets is a scientific priority \citep{natacademy_2018}.  This quest has been challenging for JWST. Two observational techniques applied to transiting systems have dominated these attempts: transit and eclipse spectroscopy.  Transit spectroscopy of rocky planets has produced only hints of spectral features, and secondary eclipses have been broadly consistent with bare rocky surfaces (\citealp{greene_2023, zieba23, xue24, mansfield24, Alderson2024, LY2023a, Kirk2024, Scarsdale2024, MoranStevenson2023, MayMacDonald2023, Lim23, Gressier2024, Cadieux2024b, Damiano2024, ducrot_2025, radica_2025, glidden_2025, espinoza_2025, gillon_2025}, reviewed by \citealp{kreidberg_2025}, and \citealp{espinoza&perrin_2025}).  Moreover, transit spectroscopy is sensitive to interference by the Transit Light Source effect \citep{rackham_2018}, notably in the TRAPPIST-1 planets (e.g., \citealp{piaulet-ghorayeb_2025}).  Thin atmospheres are possibly detected in some cases \citep{greene_2023}, and ultra-hot rocky planets can have atmospheres from magma oceans \citep{hu_2024, teske_2025}.  Nevertheless, new observational approaches to this problem are needed.  

In this paper, we develop and advocate a third observational option for temperate rocky planets, that we call color phase curves (hereafter CPC).  CPCs are based on a ratio of filter photometry at two different wavelengths observed nearly simultaneously.  For temperate planets, an optimal strategy is to choose one wavelength where a temperate planet's thermal emission is strong (e.g., 21\,$\mu$m) and another wavelength that is relatively more sensitive to the star (e.g., 12\,$\mu$m). The ratio of fluxes observed at those two wavelengths has great potential to isolate thermal emission from temperate planets, discriminate against the star, and cancel instrumental systematics. 

\cite{cowan_2007} observed several hot Jupiters at multiple wavelengths using {\it Spitzer}. \citet{stevenson_2020} proposed exploiting the infrared emission of temperate planets at long IR wavelengths, and \citet{galuzzo_2020} emphasized using JWST's MIRI photometry at 21\,$\mu$m.  Our CPC concept expands on those ideas in three ways by: 1) considering the advantages for stellar contamination (\S\ref{sec: star}), 2) using the ratio of 21- to 12.8\,$\mu$m as measured using the same detector (\S\ref{sec: instrument}), and 3) by exploring how to infer the day-to-night temperature difference without knowing the orbital inclination (\S\ref{sec: contrast}).  Although JWST's long wavelength sensitivity and low thermal background are well suited to our proposed CPC method, it has not been attempted in practice.  In that respect, JWST is arguably under-utilized for temperate exoplanets in spite of a healthy oversubscription status.  Specifically, the MIRI 21\,$\mu$m filter has yet to be used for temperate rocky exoplanets, but modeling predictions \citep{kreidberg_2016, galuzzo_2021} have been encouraging.

\subsection{Organization of This Paper}

We outline the CPC principle in \S\ref{sec: principle}, and we point out that it applies to non-transiting planets as well as transiting planets.  \S\ref{sec: star} describes how the long thermal wavelengths allow very effective discrimination against star spots and faculae, and \S\ref{sec: flares} describes the effect of stellar flares.  \S\ref{sec: instrument} discusses the potential for mitigation of instrumental systematic effects. \S\ref{sec: proxima} and \S\ref{sec: amplitudes} simulate a detection of the atmosphere of the archetype temperate non-transiting rocky exoplanet Proxima Centauri\,b.  Beyond the detection, \S\ref{sec: contrast} describes how to measure the longitudinal heat redistribution in the atmosphere, in spite of uncertainty in the inclination of the orbit, and \S\ref{sec: meridional} discusses the effect of potential meridional heat transport and also high rotational obliquities.  \S\ref{sec: sample} describes the potential for CPCs to be applied to a larger sample of nearby non-transiting planets.   \S~\ref{sec: summary} summarizes our results.

\section{Color Phase Curves (CPC)}\label{sec: cpc}

\subsection{Principle of CPC}\label{sec: principle}

Our goal in this paper is to develop the CPC principle, and illustrate how it applies to a sample of planets, including non-transiting rocky planets.  Given that non-transiting planets are more common than transiting ones, it is likely that CPC in practice will be primarily a non-transiting method.  CPC relies on the fact that the flux from a temperate exoplanet, relative to its star, increases at long IR wavelengths.  That has long been known for both both gaseous planets \citep{richardson_2003}, and rocky planets \citep{gaidos_2004, cowan_2012}, including rocky planets orbiting M-dwarfs \citep{yang_2013, cowan_2015}.

More recently, long wavelength observations with JWST have been emphasized by \citet{stevenson_2020} (the PIE method) and others \citep{limbach_2024, lustig-yaeger_2025, johnson_2026}. Our method is similar, but we place more emphasis on the Rayleigh-Jeans regime (\S\ref{sec: star}), and we discuss how the magnitude of heat transfer can be determined without knowing the orbital inclination (\S\ref{sec: contrast}). Also, relative to PIE \citep{stevenson_2020} and VPIE \citep{johnson_2026}, CPC can be viewed as making an information trade. CPC sacrifices molecular sensitivity in order to achieve the highest possible photometric precision by canceling variations in detector sensitivity (\S\ref{sec: instrument}). 

CPC requires that we know what longitudes on the planet are contributing to our measurements at every observed time.  That is easiest to do if we are confident that the planet is synchronously rotating, or at least in a lower order resonance such as 3:2 \citep{ribas_2016}. When more than one planet is present, planet-planet interactions can lead to oscillations between synchronous rotation and other spin states \citep{shakespeare_2023, revol_2024}.  In the case of the Proxima Centauri system, the 'b' and 'd' planets are not believed to be in resonance, albeit they do interact gravitationally \citep{livesey_2024}.  Our analysis is based on the case where both Proxima planets remain in synchronous rotation with their orbits.

The scientific potential of infrared phase curves for non-transiting planets has long been known \citep{seager_2009, selsis_2011, maurin_2012}.  \citet{seager_2009} pointed out that a phase curve having low amplitude can be interpreted as the consequence of longitudinal heat transport by an atmosphere, whereas a synchronously-rotating rocky planet lacking an atmosphere would be expected to have a phase variation of large amplitude (depending on the orbital inclination).  \citet{selsis_2011} described how the phase curve amplitude varies with wavelength as a consequence of the composition, pressure, and temperature of the atmosphere.  \citet{maurin_2012} investigated how multi-wavelength phase curves can be analyzed to constrain the planet's radius, albedo, and orbital inclination.  \citet{swain_2025} discussed how comparing phase curves at different wavelengths can inform us of atmospheric properties.  However, to our knowledge there is no discussion in the literature that addresses a phase curve measurement using the ratio of two wavelengths observed nearly simultaneously, especially at the long IR wavelengths we discuss in this paper.

We envision detecting and characterizing the atmospheres of temperate rocky planets by inferring their heat transport (or lack of it), following the principle of \citet{seager_2009}.  Our current formulation of CPC requires that the planet be synchronously rotating, and also we are interested in temperate planets.  Because synchronous planets are often close to their host star , they will tend to be hotter than an Earth-like temperature unless the host star has low luminosity.  Thus, temperate rocky planets orbiting M-dwarf stars are the natural targets of the CPC method.  

We choose two wavelength bands that are dominated by continuum radiation (12.8- and 21\,$\mu$m), and we avoid potentially strong absorptions such as the 15\,$\mu$m carbon dioxide band.  The reason for using continuum radiation is to maximize the amplitude of the CPC, and to avoid possible degeneracies between phase curve parameters and atmospheric composition \citep{ducrot_2025}. The 12.8\,$\mu$m filter provides a strong reference flux to enable the cancellation of variations in detector sensitivity.  Moreover, 12.8\,$\mu$m is a sufficiently long wavelength to minimize contamination from stellar activity (\S\ref{sec: star}), and avoid saturation of the detector. The choice of 21\,$\mu$m is dictated by the maximum amplitude of Proxima b's planet-to-star contrast phase curve (Table~2 of \citealp{galuzzo_2020}).  

In general, we will not know the orbital inclination of a non-transiting target planet, but in \S\ref{sec: contrast} we demonstrate that at least an approximate measurement of the longitudinal heat redistribution can be made without knowing the orbital inclination. 

\subsection{Observational Methodology}\label{sec: observe}

Transiting exoplanet observers have ample experience with phase curves, and the normal observational procedure follows \citet{knutson07} by obtaining a continuous uninterrupted time series of observations.  Not only do continuous observations densely sample orbital phases, but they also help to deal with possible drifts in instrumental sensitivity.  In principle, a phase curve could be sampled only at a small number of discrete times \citep{krick_2016}. If those samples are scheduled independently and interleaved with other observations, there is significant risk that the phase curve will be affected by uncorrectable variations in instrumental sensitivity.  Nevertheless, space-borne instrumentation is often very stable, and (for example) \citet{crossfield_2010} were able to use some discrete samples in their {\it Spitzer} phase curve of Upsilon And\,b.  Also there is evidence that MIRI filter photometry on JWST is stable, or only very slowly varying, on the long orbital time scales of temperate planets \citep{gordon_2025}.  

We envision that a CPC of a temperate rocky exoplanet would rely on discrete sampling at strategically selected times, with consecutive observations of the two photometric bands in each visit. The observations would use broad-band filter photometry to maximize the number of observed photons, and risk of saturation is reduced because stars are increasingly fainter as long IR-wavelength increases.  Radial velocity results are often sufficiently precise \citep{suarez-mascareno_2025} to allow CPC photometric observations to be scheduled near anticipated peaks and valleys in the phase curve, enhancing the precision for measurement of the amplitude.  Most important, CPC measures the integrated light of the exoplanet system (planet+star) as a {\it ratio} of the fluxes in two wavelength bands. Those two bands would be observed consecutively in the same visit, using the same detectors. Although there is no guarantee that instrumental effects will strictly cancel in the ratio, the observations can be designed to facilitate that cancellation as much as possible (e.g., \S\ref{sec: instrument}).

\section{Sources of Interference}\label{sec: interference}

\subsection{Star Spots and Faculae: Advantage of the Rayleigh-Jeans Limit}\label{sec: star}

\citet{seager_2024} point out that the nature of the Planck function facilitates transit spectroscopy at long thermal wavelengths. Emergent spectra of stars are not necessarily formed in local thermodynamic equilibrium (LTE), i.e. the source function for the emergent spectrum is not necessarily the Planck function.  Non-LTE effects are well known in strong resonance lines, and in stars having low surface gravity \citep{asplund_2005}.  Even some transitions in the thermal infrared spectra of the Sun deviate significantly from LTE \citep{carlsson_1992}.  Nevertheless, LTE is a good approximation for most of the thermal IR spectra of M-dwarf stars \citep{seager_2024}, especially in the continuum.  Moreover, line structure is weak at long IR wavelengths, so the stellar continuum dominates filter photometry.  Thus, the properties of the Planck function are central to flux variations in the stellar hosts of temperate rocky planets as observed using filter photometry. 

Because CPC uses a {\it ratio} of two wavelengths, stellar variations are potentially better behaved for CPC than \citet{seager_2024} discuss for transit spectroscopy.  The issue is to what extent variations in the stellar flux will occur due to variable contributions of spots and faculae as they form and evolve, and as the star rotates. In the Rayleigh-Jeans limit, the Planck function ($B_{\lambda}$) becomes:

\begin{equation}
B_{\lambda}(T) = 2kcT/{\lambda}^4
\end{equation}

Consider a star with photospheric temperature $T_{phot}$ and star spot temperature $T_{spot} = T_{phot} + {\delta}T$. Let the solid angle subtended by the star spot (as observed by JWST) be $\epsilon$, in units where the solid angle subtended by the JWST-facing stellar disk is unity.  Then the flux measured by JWST for the spotted star is:

\begin{equation}
F_{spotted} =  2kc[(1-\epsilon)T_{phot}+{\epsilon}T_{spot}]/{\lambda}^4,
\end{equation}

and: 

\begin{equation}
F_{unspotted} =  2kcT_{phot}/{\lambda}^4 
\end{equation}

Thus,
\begin{equation}
F_{spotted}/F_{unspotted} = 1+{\epsilon}({\delta}T/T_{phot})  
\end{equation}

Hence in the Rayleigh-Jeans limit, the flux from a spotted star relative to the flux from the same star without spots, is independent of wavelength.  (The same statement applies to faculae, or a mixture of spots and faculae, because the equations are similar.)  The CPC will be potentially contaminated by variable spot coverage, and the CPC measurements are relative in time.  Therefore, the potential contamination of the CPC varies similarly as the ratio of spotted star to unspotted star, and it is independent of wavelength. Thus, in the Rayleigh-Jeans limit, spatial variations in stellar temperature have no effect on CPCs.

The reason that wavelength cancels in $F_{spotted}/F_{unspotted}$ is because the Planck function becomes linear in temperature, which allows the wavelength dependence to be separated as a multiplicative factor, thus canceling in the flux ratio.  At short wavelengths, temperature and wavelength appear in the Planck function together in an exponential, and therefore the wavelength dependence does not factor out, and does not cancel in the flux ratio.

That being said, real measurements will not strictly be in the Rayleigh-Jeans limit, which applies at infinite wavelength.  Choosing 12.8- and 21\,$\mu$m MIRI filters, we used model atmospheres to calculate the effect for real observations. We consider the case of an M-dwarf star ($T_{eff}=3000$\,K), and a star spot temperature deficit of $\delta{T}=-200\,K$ \citep{savanov_2019}.  We use a spot area coverage consistent with the amplitude of the optical rotational light curve of Proxima Centauri\,b (Figure 7 of \citealp{wargelin_2024}).  We use Phoenix model atmospheres for both the photosphere and the star spot, and we integrate their emergent fluxes over the bandpasses of the MIRI filters and including the wavelength-dependent detector response.  The results are tabulated in Table~\ref{tab: spot_results}, and we see that in this case the difference in the flux ratio between the two wavelengths is 634 parts-per-million.  Although that is larger than the flux ratio from many temperate rocky exoplanets, it is much less than would prevail at optical wavelengths.  Also, many M-dwarfs rotate slowly \citep{newton_2018, popinchalk_2021}, and CPCs can exploit different time scales for stellar rotation versus the exoplanet's orbital period, as we illustrate in \S\ref{sec: amplitudes}. 

\begin{table}[h]
\centering
\caption {Example of the wavelength dependence of the ratio between the flux of a spotted star divided by the flux from the same star without spots.  The star is an M-dwarf with $T_{eff}=3000$\,K, with a star spot whose projected area is 0.08 of the total stellar disk's projected area.  The temperature deficit of the star spot is 200\,K.  The ratio of spotted-to-unspotted flux is almost equal at the two MIRI filter wavelengths (integrated over the bandpasses); the difference in the ratio is 634 ppm.} 
\begin{tabular}{ll}
Wavelength ($\mu$m) &  $F_{spotted}/F_{unspotted}$   \\
\hline 
\hline
12.8 & 0.990855 \\ 
21 & 0.990221 \\
\hline
\end{tabular} 
\label{tab: spot_results}
\end{table}

\subsection{Stellar Flares}\label{sec: flares}

Stellar exoplanet hosts can exhibit flares at infrared wavelengths.  For example, \citet{seager_2009} encountered a flare in their exploratory observations of GJ\,876, and \citet{gillon_2025} found multiple flares on TRAPPIST-1.  The advantages of the Rayleigh-Jeans regime are not relevant to flares, because their mid-IR emission is not thermal \citep{kaufmann_2004}.   In the case of Proxima Centauri, flares occur during about 7\% of observations \citep{vida_2019}.  Fortunately, flares are easily recognized by their sharp rise time \citep{seager_2009, yang_2025}.  To the extent that they can be detected in CPC data, they can be either corrected by model fitting \citep{gillon_2025}, or zero-weighted using digital filters.  However, it is possible that a population of low-amplitude flares may be more difficult to identify and correct than are large events. This is a topic that needs additional investigation.

\subsection{The Instrument}\label{sec: instrument}

In these simulations we have adopted photon-limited performance.  By photon-limited we specifically mean that the total noise comprises only photon noise from the host star, but also accounting for noise in subtraction of the thermal background (i.e., telescope and zodiacal emission).  Even at the long wavelengths we envision, the candidate stars for CPCs are sufficiently bright that their photon noise dominates over detector read noise. Yet they are fainter at these long wavelengths than in the optical and near-IR.  Consequently, a photon-limited signal-to-noise ratio is less in the thermal-IR than at brighter wavelengths, and in principle easier to attain.  

In addition to photon noise, time series observations of transiting planets often exhibit electronic red noise from the infrared detectors. Such effects were prominent using {\it Spitzer} (e.g., \citealp{deming_2006}), but have improved in JWST observations \citep{kreidberg_2025}. \citet{zieba_2026} attained a total noise that was 3\% to 7\% above the photon noise for MIRI/LRS spectroscopic observations.  For MIRI filter photometry, electronic detector noise has been investigated on time scales shorter than a single integration \citep{xue_2025}, and also on time scales of hours \citep{holmberg_2026}, and many days \citep{gordon_2025}.  After correcting detector noise using a Gaussian process, \citet{holmberg_2026} obtained a white noise component only 20\% above the photon noise in their filter photometry.  Moreover, \citet{august_2025} and \citet{meier_valdes_2025} found that the precision of their photometry results binned down with time close to the theoretical square-root dependence, at least on the time scale of several hours.


Observing a color, i.e. two wavelengths, can in principle facilitate the mitigation of instrumental systematic errors in the detectors. \citet{cowan_2007} measured phase curves of several hot Jupiters in multiple wavelength bands using {\it Spitzer}. They found that the precision for colors was significantly worse than the photon noise, but Spitzer used different detectors at different wavelengths.  In the JWST era, we anticipate that the ratio between two wavelengths can be measured more precisely than photometry in a single wavelength, because the same detector - and to a large degree even the same pixels - can be used for both wavelengths (e.g., JWST/MIRI has one detector array for filter imaging). Different wavelengths will not illuminate the detector in the same way, because of wavelength-dependent diffraction.  That mismatch can be at least partially compensated by dithering.  Nevertheless, there will probably be residual effects wherein detector sensitivity does not cancel perfectly between the two wavelengths.  Hence we must consider how those effects will impact the CPC, and we consider both short term (hours) and long term (days) variations in detector response. 


Short-term changes in detector response include effects of charge-trapping and persistence.  However, these effects in JWST/MIRI photometry stabilize quickly for bright sources, as recently demonstrated by \citet{connors_2025}. Proxima is even brighter than HD\,260655 that caused the MIRI detector ramp to flatten rapidly (Figure~5 of \citealp{connors_2025}). The rapid flattening of detector ramps for bright sources like Proxima is consistent with the experience of Spitzer observers, who sometimes observed a bright extended source immediately prior to a transiting system in order to 'pre-flash' and stabilize the detector \citep{knutson_2011}. Long-term changes could also occur, investigated by \citet{gordon_2025}, and are greater at 21\,$\mu$m than at 12.8\,$\mu$m. Fortunately, \citet{gordon_2025} found that long-term changes decay exponentially, and we are now far from their starting epoch (Julian date $t_{0}=2459720$).  From their fitted curves, we expect a change in relative detector response (21\, vs. 12.8\,$\mu$m) that is insignificant for the purpose of CPC observations.

Finally, we point out that photon-limited performance can be calculated accurately, because the brightness of planet-hosting stars is usually well known, and the sensitivity of JWST is well calibrated \citep{gordon_2025}.  Whereas red noise from the detectors is much more difficult to predict accurately.  It is logical to adopt photon-limited performance as a baseline, and then to point out (e.g., Section~\ref{sec: amplitudes}) how the results are affected if additional noise is present.

\section{An Example: Detecting the Atmosphere of Proxima Centauri\,{b}}\label{sec: proxima}

To illustrate how CPC would work in practice, we use the example of Proxima Centauri. This system has two planets: b and d.  Proxima b \citep{anglada_2016} is within the star's habitable zone and temperate ($T\sim 218$\,K),  and rocky with 90\% probability \citep{bixel_2017} - especially given the recently-inferred orbital inclination of 47-degrees \citep{klein_2021, suarez-mascareno_2025}. An orbital eccentricity $e<0.1$ \citep{suarez-mascareno_2025} makes it likely to rotate synchronously.  It does not transit \citep{jenkins_2019, gilbert_2021}.  Proxima d \citep{suarez-mascareno_2020, suarez-mascareno_2025} is smaller in mass, radius, and orbital distance, also rocky, and likely synchronous.  Table~\ref{tab: params} summarizes the basic properties of Proxima's planets, but those numbers are for context only - we calculated emergent fluxes for specific models as described in \S\ref{sec: amplitudes}.   A third planet ('c') has been reported \citep{damasso_2020}, but is still uncertain \citep{artigau_2022}, so we do not include it here.  Table~\ref{tab: obs_summary} summarizes the parameters of a JWST observational program that we envision and disccuss below.

\begin{table*}[h]
\centering
\caption {Parameters of the Proxima system.  Period is orbital for the planets and rotational for the star, all in days. Radii are in Suns for the star, and Earths for the planets. Sources are [1: \citet{boyajian_2012}], 
[2: \citet{suarez-mascareno_2020}], [3: \citet{mann_2015}], [4: calculated in this work using the empirical mass-radius relation from \citet{muller_2024}].} 
\begin{tabular}{llll}
   &  Star &  'b' planet  &  'd' planet \\
\hline 
\hline
Temperature &  $3050\pm100$\,K [1] &  $218\pm18$\,K [2] &  $282\pm23$\,K  [2] \\ 
Mass &  $0.122\pm0.002$  [3] &  $1.44\pm0.21$ [2] &  $0.357\pm0.072$ [2] \\ 
Radius &  $0.141\pm0.021$ [1] &  1.12 [4]  &   0.77 [4]   \\ 
Period &  $83.2\pm1.6$ [2]  &  $11.18465\pm0.00053$ [2]   &  $5.12338\pm0.00035$ [2]    \\ 
\hline
\end{tabular} 
\label{tab: params}
\end{table*}

\begin{table}[h]
\centering
\caption {Summary of a CPC observing program for the Proxima Centauri system. The exposure time per visit is equally divided between 12.8- and 21\,$\mu$m.  The precision per visit applies to the ratio of 21- to 12.8\,$\mu$m, and is calculated assuming photon-noise-limited observations.} 
\begin{tabular}{ll}
\hline 
\hline
Exposure time per visit &  3.6 hours \\ 
Number of visits &  25   \\ 
Total exposure time, all visits & 90 hours \\
Precision per visit &  8~ppm  \\  
\hline
\end{tabular} 
\label{tab: obs_summary}
\end{table}

Based on the stellar and orbital parameters in Table~\ref{tab: params}, we calculated CPCs for the Proxima system ('b'+'d' planets) using several models.  First, we used the 'Toy Climate Model' from \citet{kreidberg_2016}, applying that model also to the 'd' planet by scaling the 'b' phase curve to the temperature and radius of 'd'.  For the 'b' planet, we also include the results from the GCM calculation of \citet{galuzzo_2020}.  Those authors do not tabulate their full phase curves, but their Figure~8 appears very sinusoidal, with a shape similar to the toy model from \citet{kreidberg_2016}.  We therefore include the GCM result by scaling the toy model to the GCM amplitudes and orbital inclination, using Tables~2 and 3 in \citet{galuzzo_2020} to mimic a GCM version of the CPC for planet 'b'.

\begin{figure*}[h]
\centering
\includegraphics[width=\textwidth]{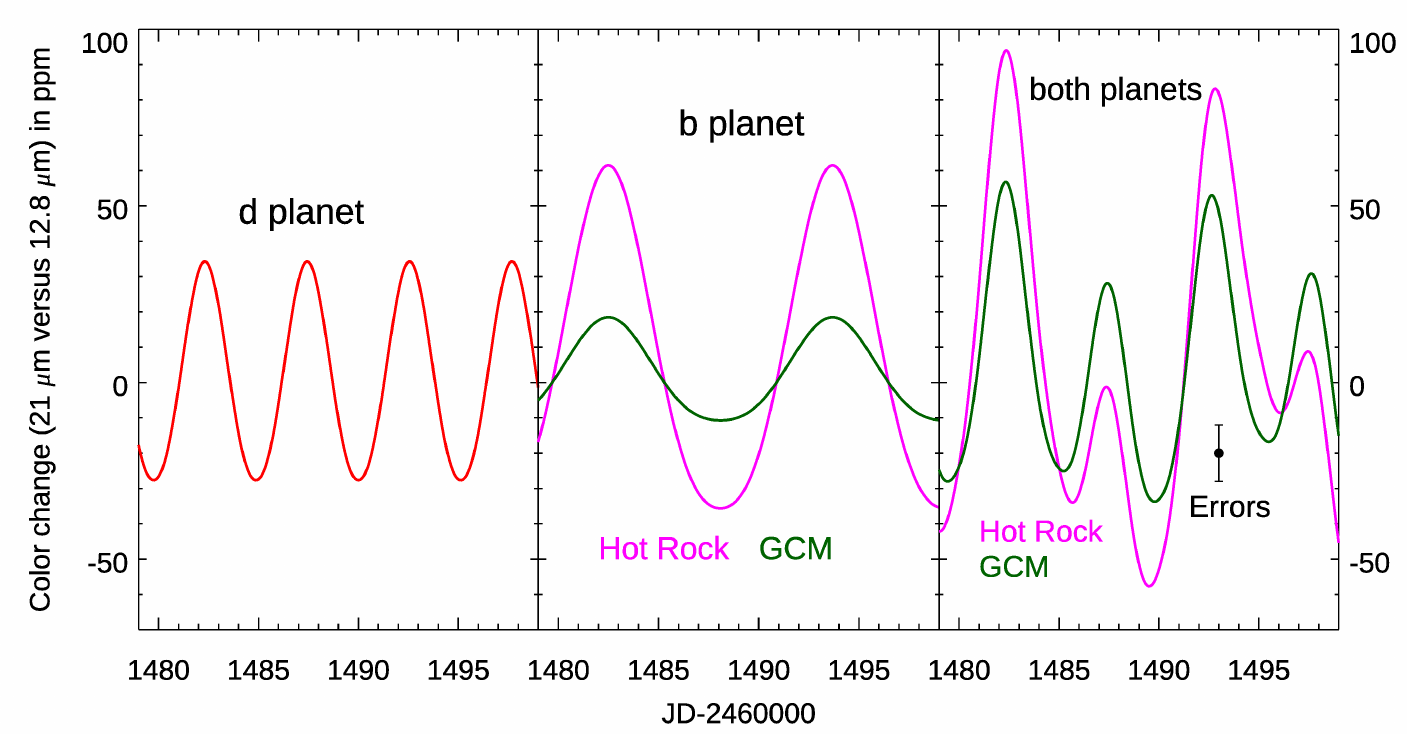}
\caption{Color phase curves (CPCs) for planets 'b' and 'd' in the Proxima Centauri system.  The CPC (Y-axis) is the ratio of the flux in the 21\,$\mu$m to 12.8\,$\mu$m JWST/MIRI filters, normalized to the stellar flux, with the median value subtracted from the normalized ratio. The time period (X-axis) is a random 20-day interval during JWST's 2027 visibility window.  The leftmost panel shows the 'd' planet only.  Two cases are shown for the 'b' planet (middle panel):  using a hot rock model, and a GCM (see text).  The rightmost panel shows the sum of 'b' and 'd' for the two 'b' cases. Planet 'd' is modeled as a hot rock in both cases. }
\label{fig: curves1}
\end{figure*}

CPCs for the sum of both planets at an orbital inclination of 47-degrees \citep{suarez-mascareno_2020} are shown in Figure~\ref{fig: curves1}. The toy models for 'b' and 'd' in this case are for zero heat redistribution, and a Bond albedo of 0.1, = dark, hot rocks.  The GCM amplitude from \citet{galuzzo_2020} corresponds to a Bond albedo near 0.3, = Earth-like.  \citet{galuzzo_2020} do not quote the degree of bolometric heat redistribution produced by their GCM, but from the contrast in their phase curve (maximum versus minimum), we estimate the redistribution to be approximately 40\% (on a scale where complete redistribution in longitude is 100\%).

Figure~\ref{fig: curves1} includes the 'd' planet as a hot rock, added to both the toy model and the GCM for planet 'b'.  We used a Phoenix model for the star, and we integrated the planetary and stellar models over the bandpasses of the MIRI 12.8- and 21\,$\mu$m filters, including variations in the detector response with wavelength. The CPCs for the two planets come in and out of phase with each other, with a total peak-to-peak variation (both planets) of $\sim 160$\,ppm for the toy model applied to 'b', and $\sim 100$\,ppm using the GCM model for b ('d' is a hot rock in both cases).  We envision a series of observations (described in \S\ref{sec: amplitudes}, and summarized in Table~\ref{tab: obs_summary}), wherein the total exposure time (12.8\,$\mu$m + 21\,$\mu$m) equals 3.6 hours per each of 25 visits, and we use the MIRI Exposure Time Calculator to calculate the photon-limited noise for each visit (= 8\,parts-per-million).  

\begin{figure*}[h]
\centering
\includegraphics[width=\textwidth]{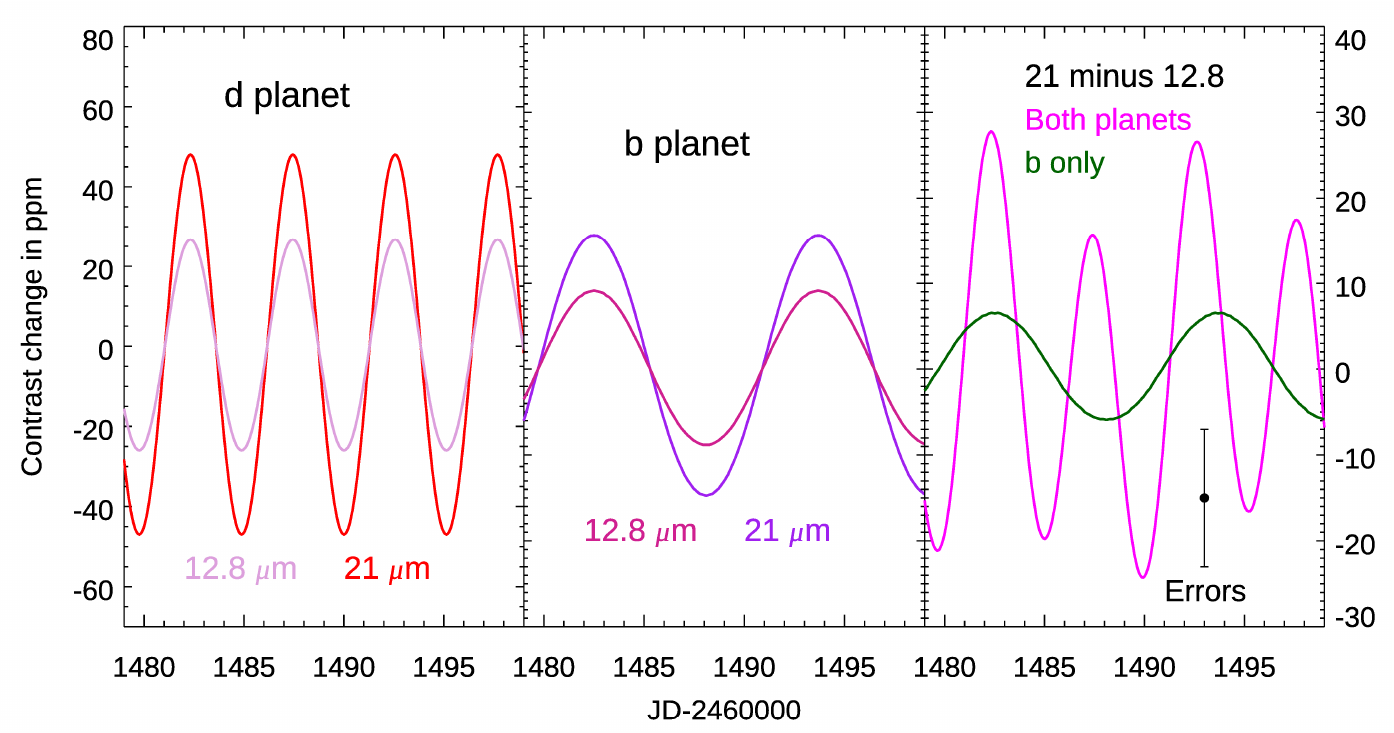}
\caption{Fluxes relative to the host star, and color phase curves (CPCs) for planets 'b' and 'd' in the Proxima Centauri system. The left Y-axis scale applies to only the leftmost panel; the right Y-axis scale applies to the center and rightmost panel. The time period (X-axis) is a random 20-day interval during JWST's 2027 visibility window.  The left and center panels plot fluxes relative to the star at both 12.8\,$\mu$m and 21\,$\mu$m.  The right panel plots the CPC for the sum of both planets, and for 'b' separately. Both planets are represented by the two-column model of \citet{lincowski_2023} (see text for description).}
\label{fig: curves2}
\end{figure*}

We explored additional models for both the 'b' and 'd' planets, using the two-column formulation described by \citet{lincowski_2023}.  We emphasize that these models are not intended as a prediction for this planetary system based on first principles; rather, they are plausible models to test the detectability of this system.  The two-column method adopts day and night-side atmospheric columns, and transport between them is calculated for each vertical layer using the 3D primitive equations.  A key advantage of this method is that radiative transfer can be treated more completely than is often possible using GCMs.  Radiative transfer was calculated using the Spectral Mapping Atmospheric Radiative Transfer code (SMART).  SMART uses a multi-stream, multi-scattering algorithm at high spectral resolution \citep{meadows_2018, meadows_1996}. The hemispherically-averaged day side column does not spatially resolve the narrow substellar region \citep{gillon_2025}, hence these models probably underestimate the phase curve amplitudes and thereby provide conservative estimates of detectability.

 Our two-column model for the 'b' planet is an Earth-like atmosphere with the volume mixing ratio of carbon dioxide equal to 280 parts-per-million. Figure~\ref{fig: curves2} shows the CPC for that model; it has an albedo of 0.3, and 35\% of the incident stellar energy transported to the night side. The 'b' amplitude from the two-column model is less than the GCM model due to the hemispherical averaging mentioned above (also, see \S~\ref{sec: amplitudes}), but the Bond albedo and the heat circulation are similar.  Figure~\ref{fig: curves2} also uses a two-column model for the 'd' planet.  That 'd' model is a "Dune"-like planet having Earth-like atmospheric gases except a 10\% relative surface humidity, no clouds, and albedo = 0.1.  It has no transport to the night side.

\section{Measuring the Phase Curve Amplitude}\label{sec: amplitudes}

The first inference that can be made from CPC data is a retrieval of the separate amplitudes of the CPC for each exoplanet in the system that contributes significantly.  We have simulated this process for the Proxima system's CPCs shown in Figures~\ref{fig: curves1} and \ref{fig: curves2}, sampling the CPCs using the observational parameters summarized in Table~\ref{tab: obs_summary}. The radial velocity data \citep{suarez-mascareno_2025} are sufficiently precise to enable observations to be selectively scheduled at the times where peaks and valleys are expected to occur in the total CPC.  We simulated 50,000 sets of observational data, all based on the same targeted observational windows, but with the exact time varying randomly within each $\pm0.5$\,day window, to mimic the scheduling constraints of real observations.  We create the data samples from the modeled CPCs, summing both the 'b' and 'd' planets, and adding Gaussian random noise to each observation with a standard deviation of 8 ppm.  We also varied the planetary orbital inclinations (taken to be coplanar) when generating each set of data, and also their orbital periods, using Gaussian distributions corresponding to the current observational uncertainties (Table~\ref{tab: params}). In addition to the total CPC due to planets 'b' and 'd', we add stellar variation that is modeled as sinusoidal at the 83.2-day rotational period, and having a peak-to-peak amplitude of 634 ppm in the 21- versus 12.8\,$\mu$m color (\S\ref{sec: star}, and Table~\ref{tab: spot_results}).  

For each set of simulated data, we solve for the amplitudes of the CPCs for 'b' and 'd', and the amplitude of stellar variation, using multi-variate linear regression. In real data, it is possible that the stellar variation could be at least partially inferred from the absolute fluxes (Table~\ref{tab: spot_results}), but our regression solutions do not attempt that. The regression solutions assume that the planetary orbital periods are always equal to the current best-estimate values (although we vary them in the simulated data), and the regression solves for their amplitudes.  Since the regression does not know the orbital inclination, it adopts 47-degrees \citep{klein_2021} in all cases; consequently the retrieved amplitudes are degenerate with the variations in inclination that are included in different realizations of the data.

\begin{figure*}[h]
\centering
\includegraphics[width=\textwidth]{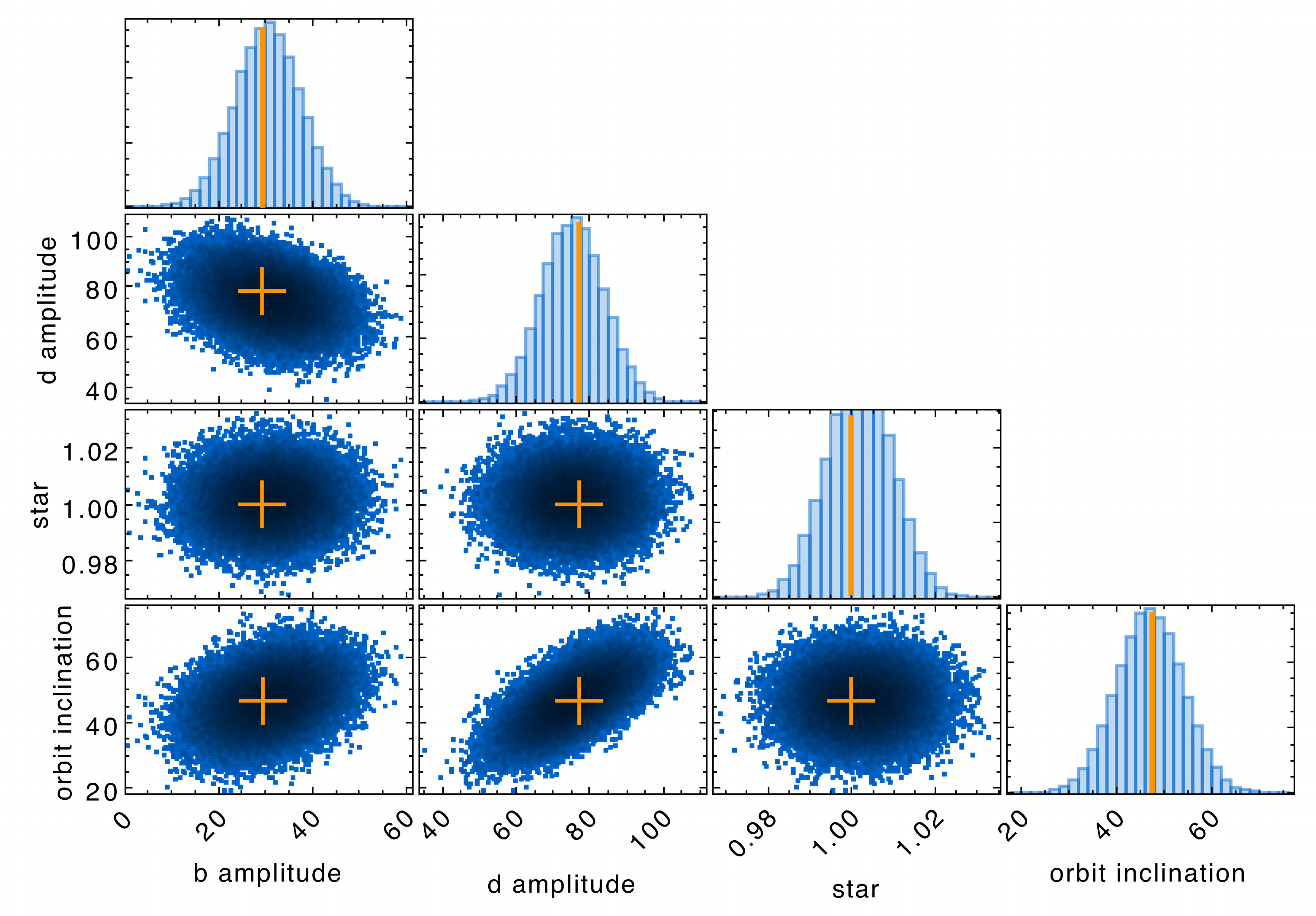}
\caption{Corner plot for multi-variate linear regression results from 50,000 sets of simulated CPC data for the Proxima system. Each data set is based on 25 observations of the CPCs shown in Figure~\ref{fig: curves1}, with 8~ppm error bars (see Table~\ref{tab: obs_summary}).  The planets are taken to be coplanar, and the regression solves for the CPC amplitudes of planets 'b' and 'd', as well as the amplitude of stellar variation.  The orbital inclination is varied for each simulated dataset, with the amplitude of variations corresponding to the current observational uncertainties (Table~\ref{tab: params}). The orange crosses and lines mark the values used to generate the simulated data.}
\label{fig: corner1}
\end{figure*}

Figure~\ref{fig: corner1} shows the corner plot for the data simulations corresponding to Figure~\ref{fig: curves1}. The amplitude of the 'd' curve is degenerate with the orbital inclination as expected.  The amplitude for 'b' however, is less degenerate with inclination.  We attribute that to the nature of the day- versus night-side temperature contrast that is much less for 'b' because of the presence of an atmosphere with heat redistribution. The retrieved median amplitude of 'd' is $75.4\pm8.6$\,ppm, and for 'b' is $30.5\pm7.0$\,ppm, where the uncertainties include degeneracy with the orbital inclination. The input values are 77 and 29\,ppm, respectively. If the precision is degraded from the photon-limited case by 20\% \citep{holmberg_2026}, the detections are still secure at 7.8$\sigma$ and 3.6$\sigma$ respectively.  Had we used the toy model for 'b' (a ``hot rock") with no redistribution of heat, the amplitude would be 114\,ppm - a high significance detection, even if the precision is degraded from the photon-limited case.  We conclude that summed CPCs for two planets in the Proxima system can be successfully decomposed into their individual amplitudes, if the 8\,ppm precision that we project for the color measurements can be achieved.  Also, Figure~\ref{fig: corner1} shows essentially no degeneracy between the planetary amplitudes and the stellar variation, and that is not surprising given their very different time scales (Table~\ref{tab: params}). 

\begin{figure*}[h]
\centering
\includegraphics[width=\textwidth]{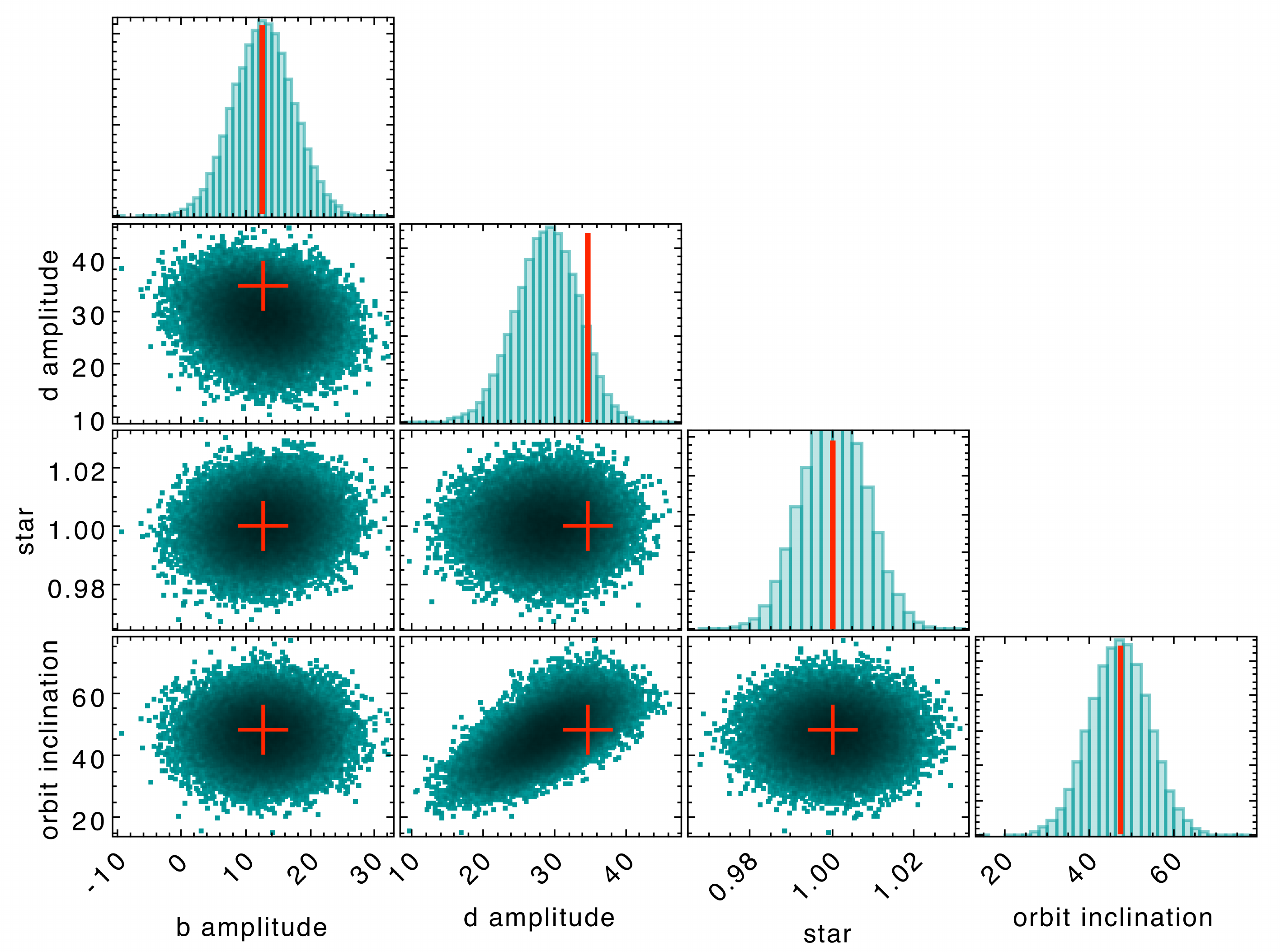}
\caption{Corner plot for the sum of Proxima 'd', and 'b', using our two-column models for both planets.  The red crosses and lines mark the values used to generate the simulated data.  The observational parameters for the simulated data are summarized in Table~\ref{tab: obs_summary}. }
\label{fig: corner2}
\end{figure*}

Figure~\ref{fig: corner2} shows the retrieved CPC amplitudes for the case where both planets are represented by our two-column model, as in Figure~\ref{fig: curves2}.  The retrieved amplitudes for 'd' and 'b' are $29.0\pm4.5$\,ppm and $12.7\pm4.7$\,ppm, both in good agreement with their input values of 34.7 and 12.5\,ppm, respectively.  Compared to Figure~\ref{fig: corner1}, the 'd' amplitude is still degenerate with the orbital inclination, but the retrieved median value agrees with the 34.7\,ppm input to $1.3\sigma$. The retrieved 'b' amplitude ($12.7\pm4.7$\,ppm), is a marginal detection but in good agreement with the 12.5\,ppm input value (see Figure~\ref{fig: curves2}). If the precision is degraded by 20\% over the photon-limited case \citep{holmberg_2026}, then 'd' is still securely detected (5.7$\sigma$), but 'b' becomes only 2.2$\sigma$, not a detection.  Regardless of the statistical significance of the 'b' amplitude, the comparison with 'd' would be very significant evidence for an atmosphere that is causing heat redistribution in the 'b' planet (otherwise the 'b' amplitude would be larger).  That comparison assumes that the two planets are coplanar, as in the TRAPPIST-1 system \citep{gillon_2017}.  An alternative would be a large difference in orbital inclination between the two planets.  But a large inclination difference could be dynamically unstable \citep{livesey_2024}, and in any case would be a significant result in its own right.

\section{Day-to-Night Temperature Contrast}\label{sec: contrast}

We envision CPCs detecting and characterizing rocky exoplanet atmospheres not by detecting spectral features, but by sensing the day-to-night temperature contrast, and inferring the requisite magnitude of heat transport \citep{seager_2009}. By comparing the observed CPC amplitude with models, the magnitude of heat transport can in principle be determined.  CPC amplitude varies with the albedo of the planet, as well as with the planetary radius and orbital inclination. Models such as our two-column model do not necessarily use heat transport and albedo as input parameters.  Rather, those quantities are often determined self-consistently by the physics of the model, given starting boundary conditions such as the bulk composition of the atmosphere and the stellar irradiance. Hence the first goal of observations is to infer the magnitude of the day-to-night variation in IR brightness temperature, independent of the planet's radius and orbital inclination.  With that temperature contrast determined, in-depth physical models can be invoked to infer the magnitude of heat transport.

\begin{figure}[h]
\centering
\includegraphics[width=3.0in]{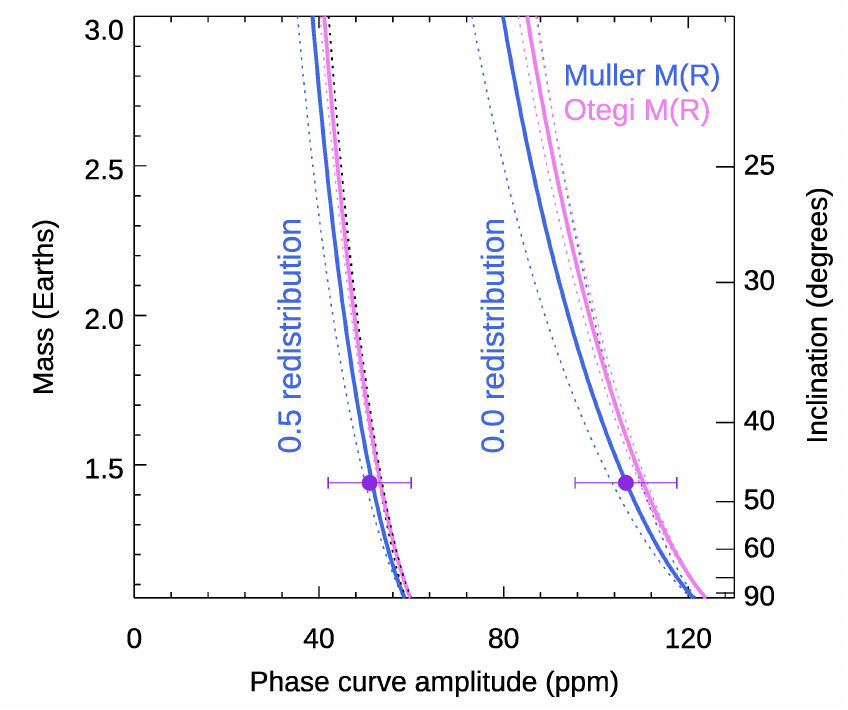}
\caption{Possible mass and orbital inclination of Proxima Centauri\,b as a function of the amplitude of the CPC (21- vs. 12.8\,$\mu$m), where each curve is constrained by a mass radius relation.  The phase curve amplitudes (X-axis) are based on the toy models from \citet{kreidberg_2016} (see text). We adopt M(R) relations from \citet{otegi_2020} and \citet{muller_2024} in order to show that the result is not critically dependent on the exact form of M(R). Dotted lines show the $\pm1\sigma$ error ranges due to imprecision in M(R). Because the curves for different values of heat redistribution are nearly vertical, an observed CPC amplitude will intersect only a small range of heat redistribution values (see text). \label{fig: MRplot}  }
\end{figure}

We here demonstrate that day-to-night temperature contrast, and consequently the magnitude of heat redistribution, can be determined without knowing the orbital inclination, by invoking a mass-radius relation.  For this purpose, we use the toy models of \citet{kreidberg_2016}, because they can specify the degree of heat redistribution as an input parameter. We determine the magnitude of heat redistribution by comparing the measured phase curve amplitude to models that are expressed in a quasi-inclination-independent form (Figure~\ref{fig: MRplot}). Prior to implementing that quasi-inclination-independent form, we first consider the plausibility of the determination by invoking some general principles.

\subsection{Plausibility of Measuring Heat Transport}

First, we consider the number of physical variables that contribute to an observed CPC for the 2-planet Proxima system.  Those are: $R_{b}$, $R_{d}$, $i_{b}$, $i_{d}$, $\epsilon_{b}$, $\epsilon_{d}$, $\alpha_{b}$, and $\alpha_{d}$, where $R$ is radius, $i$ is orbital inclination, $\epsilon$ is the fraction of incident energy transported to the night side (and re-emitted there), and $\alpha$ is the Bond albedo.  Subscripts $b$ and $d$ refer to the two planets we consider in this system.  At this point we have eight unknown variables, but only 4 observables: CPC amplitudes for each planet, and the radial velocity measurements of $M\sin{i}$ for each planet.  By adopting a mass-radius relation (equivalent to an inclination-radius relation), and assuming that the orbital inclinations are the same (co-planar), we can reduce the number of unknowns to five, so the problem is still under-constrained.  However, the under-constrainment is not extreme, and models can be used to reduce the number of unknowns to equal the number of observables, as we now describe using a quasi-inclination-independent formulation.

\subsection{A Quasi-inclination-independent Formulation}

The inclination-independent form is achieved as follows: The RV data ($M\sin{i}$) imply orbital inclination as a function of possible mass, and a mass-radius relation \citep{otegi_2020, muller_2024} gives radii.  For each value of heat redistribution, the models from \citet{kreidberg_2016} thereby predict phase curve amplitudes versus mass. Figure~\ref{fig: MRplot} plots those mass-amplitude relations, but with mass as the dependent variable. We use two versions of the mass-radius relation in order to demonstrate that the results are not strongly sensitive to the version of the mass-radius relation.  The curves in Figure~\ref{fig: MRplot} are almost vertical because: 1) with increasing mass and radius, the phase curve amplitude in IR color increases, but 2) increasing mass requires smaller orbital inclinations and that works in the opposite direction.  

For a given albedo, a measured phase curve amplitude (X-value on Figure~\ref{fig: MRplot}), intersects only a small range of model curves that define the heat redistribution.  In this form, the under-constrained aspect of the problem appears as the albedo. Fortunately, in the Rayleigh-Jeans limit the (Bond) albedo affects the phase curve amplitude only weakly.  That occurs because the infrared flux is only linear (not exponential) versus temperature, and temperature is proportional to the fourth-root of the absorbed energy.  Also, self-consistent physical models can be invoked to simultaneously determine consistency between the heat transfer and the albedo.  In other words, the physical models connect the albedo and the heat transfer efficiency and provide the final relation that reduces the number of unknowns to equal the number of observables. Indeed, some possibilities can be constrained even without full modeling: for example the zero-redistribution curves on Figure~\ref{fig: MRplot} have sufficiently large amplitudes to require albedos not greatly exceeding 0.3.

\subsection{Meridional and Zonal Circulation, and Rotational Obliquity}\label{sec: meridional}

\subsubsection{Emulating Variations in Atmospheric Circulation}

The atmospheric circulation of rotating planets is primarily zonal, due to the Coriolis effect \citep{pierrehumbert_2010}.  Nevertheless, some degree of meridional circulation is possible and could potentially affect CPCs.  \citet{schwartz_2015} found that meridional circulation produced a negligible effect on the observed phase curves of hot Jupiters.  Our two column model uses two temperature profiles, one for the day side hemisphere and one for the night side. In effect, it assumes perfect meridional distribution of heat as well as perfect zonal redistribution on each hemisphere.  The models from \cite{kreidberg_2016} use no meridional circulation, and thus our results effectively cover the range of possibilities from zero to maximum meridional circulation.  Nevertheless, we have explored the effect of reducing the meridional and zonal circulation that is inherent in our two column model.

We emulate the effect of reducing the heat redistribution by weighting the radiance emitted from each position on the day side hemisphere by a factor of $4(\cos{\delta\theta})/\pi$, where $\delta\theta$ is an angular distance from the sub-stellar point.  When reducing only meridional redistribution, $\delta\theta$ is taken to be the difference in latitude.  When reducing both zonal and meridional redistribution, $\delta\theta$ is taken to be the total angular distance from the sub-stellar point.  The factor of $4/\pi$ is needed to keep the total emitted flux from the hemisphere constant. This is an approximate treatment because the weighting factor is applied to the emergent radiance using a single representative temperature structure, as a function of angle, rather than with a 3D GCM.

We find that reducing the meridional redistribution of Proxima b to effectively zero has a negligible effect on the CPC (less than 1\,ppm).  That happens because its orbital inclination (47 degrees) is almost exactly mid-way between pole-on and equator-on.  In the pole-on case, the CPC amplitude is reduced when zeroing the meridional redistribution, whereas the CPC amplitude is increased in the equator-on case.  Proxima b's intermediate orbital inclination produces almost no effect on the CPC.  We did not explore the effect of zeroing Proxima b's zonal redistribution because that would correspond to a no-atmosphere case, and a fundamental postulate of this paper is that Proxima b has an atmosphere.

In the case of Proxima d, we explored the case of reducing both the meridional and zonal redistribution, resulting in an ``eyeball" planet having a peak radiance at the sub-stellar point.  That increases the CPC amplitude to 175\,ppm.  While that would be an easy detection for JWST in the photon limit, we also tested whether it might bias or distort the process of simultaneously extracting a CPC amplitude for Proxima b.  We repeated the data simulations of \S\ref{sec: amplitudes} (Figure~\ref{fig: corner2}), and we obtained $167\pm17$\,ppm for the CPC amplitude of 'd', and $12.7\pm4.7$\,ppm for 'b'.  Those extracted values are consistent with the input values of 175 and 12.5\,ppm, respectively.  We conclude that a CPC-based detection of the atmosphere of Proxima b is not likely to be strongly affected by plausible variations in the nature of atmospheric circulation on both planets.

\subsubsection{Rotational Obliquity}

Our simulations adopt the classic case of zero rotational obliquity and synchronous rotation.  \citet{hammond&komacek_2024} used a GCM to study the effect of varying rotational obliquity for TRAPPIST-1e, inferred to be synchronously rotating.  In the limit of low atmospheric CO$_2$ column density (i.e., a warm rock), their Figure~8 shows that the phase curve is insensitive to rotational obliquity.  For thick CO$_2$ atmospheres, the phase curve becomes more complex, in the sense that shorter periods appear at high obliquities, but the amplitude is not strongly damped.  We conclude that possible high rotational obliquities are not an impediment to CPC measurements, but it might become necessary to consider rotational obliquity in the interpretation of measured CPCs (e.g., \citealp{gaidos_2004, cowan_2012}).

\section{The Larger Potential Sample}\label{sec: sample}

\begin{figure}[h]
\centering
\includegraphics[width=3.5in]{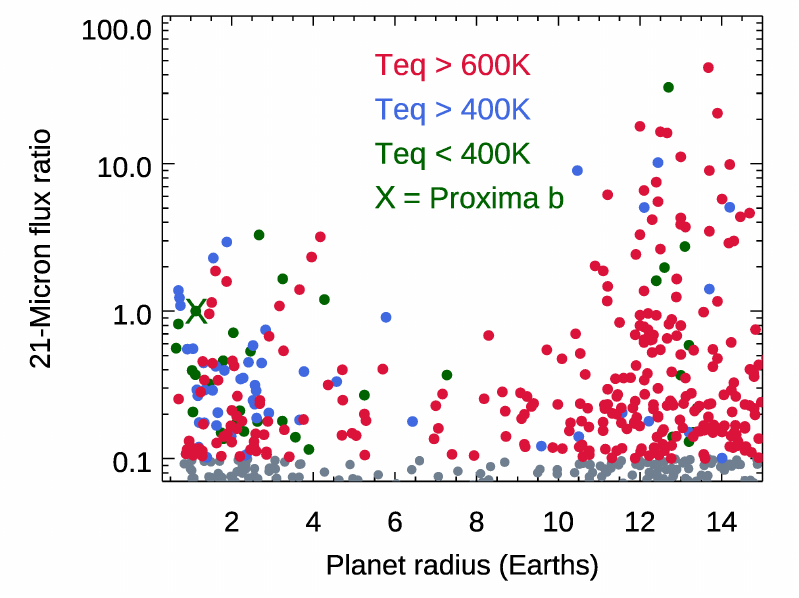}
\caption{Predicted 21\,$\mu$m flux for both transiting and non-transiting exoplanets, divided by the predicted flux from Proxima~b, where the predicted fluxes are based on radii, distances, and equilibrium temperatures from NexSci\footnote{\url{https://exoplanetarchive.ipac.caltech.edu/cgi-bin/TblView/nph-tblView?app=ExoTbls&config=PSCompPars}}. Points on the plot are color-coded by equilibrium temperature, and Proxima~b is marked by the green X at the mid- to lower-left. Gray points at the bottom fall below the flux cutoff that we consider (see text). \label{fig: flux}. }
\end{figure}

Although we have emphasized Proxima~b as the archetype of temperate rocky exoplanets, we emphasize that CPCs could be productive to detect and characterize atmospheres on a wide variety of exoplanets, including non-transiting ones, and including gaseous planets as well as rocky ones.  Figure~\ref{fig: flux} shows the predicted 21\,$\mu$m fluxes received at Earth for all exoplanets discovered by either transits or radial velocities, as a ratio to the flux predicted for Proxima~b.  Although Proxima~b is the closest exoplanet, others can have greater fluxes if they have larger radii, and/or they are warmer. Also, orbital inclinations of most systems will be greater than Proxima's 47-degrees, facilitating the inference of heat redistribution as per Figure~\ref{fig: MRplot}.

Figure~\ref{fig: flux} color-codes planets with 21\,$\mu$m fluxes exceeding 10\% of Proxima~b.  Although some of our models for Proxima~b are only marginally detectable (e.g., Figure~\ref{fig: corner2}), it is nevertheless reasonable to expect useful detections of some planets with lower predicted fluxes. The reasons are that the primary criterion is to be able to rule out the 'hot rock' case, and infer the existence of an atmosphere, even if that atmosphere does not produce a significant CPC amplitude (e.g., a planet with a spatially uniform temperature).  The 'hot rock' cases will produce maximal CPC amplitudes.  Also, other planetary systems may have orbital inclinations that are more favorable than Proxima's 47-degrees.  Finally, we note that the signal-to-noise (S/N) for the photon-limited color measurements that we envision is proportional to the square root of flux, so that S/N does not degrade rapidly as flux decreases.  Based on all of those considerations, we illustrate planets having flux ratios exceeding 0.1 in Figure~\ref{fig: flux}.  Those planets include many hot and warm gaseous planets (Jupiters to Neptunes) as well as additional temperate rocky planets.  Considering the temperate range, we find 5 rocky planets (other than Proxima 'b' and 'd').  Those 5 are: Barnard\,e \citep{basant_2025}, GJ411\,b \citep{diaz_2019}, Ross\,128\,b \citep{bonfils_2018},  YZ\,Ceti\,d \citep{astudillo-defru_2017}, and Teegarden\,b \citep{zechmeister_2019}.  They were all detected using radial velocity, and are not known to transit.  They all orbit M-dwarf stars, and the activity of some host stars (especially flares) could be problematic for atmospheric detection and even for existence of planetary atmospheres \citep{garcia-sage_2017}.  However, not all M-dwarfs are active (e.g., Ross\,128 is quiet, \citealp{bonfils_2018}), and the long-wave photometric capability of JWST gives us a powerful potential probe of these systems.  One caveat is that the CPC analysis must know how the Earth-facing hemisphere of the planet varies with time.  That is facilitated if we have high confidence that the planet's rotation is synchronous with the orbit. Finally, we caution that the JWST detector ramps may take much longer to stabilize for sources significantly fainter than Proxima (\citealp{connors_2025} and \S\ref{sec: instrument}). 

We are reminded of early work that attempted to detect the atmospheres of close-in radial velocity planets prior to the discovery of transits.  For example, both \citet{charbonneau_1999} and \citet{collier_cameron_1999} searched for reflected light from the non-transiting hot Jupiter Tau~Bootis\,b. That pioneering work was quickly supplanted by the enormous productivity of transit observations. Meanwhile, observational capabilities have continued to improve on many fronts. Radial velocity observers are detecting numerous rocky planets, and most of them do not transit.  Consequently, we suggest that there may be value in re-thinking alternate methods, and color phase curves are a potentially valuable option.  

\section{Summary}\label{sec: summary}

In this paper, we develop and advocate an alternative method to detect and characterize exoplanetary atmospheres, emphasizing the importance for temperate rocky exoplanets.  We have made these points:

\begin{itemize}

\item  In \S\ref{sec: intro} we point out that methods to measure the atmospheres of rocky exoplanets have focused on two methods: transit spectroscopy (e.g., the TRAPPIST-1 system), and secondary eclipse photometry.  The challenges that these methods have encountered motivates additional approaches. For temperate planets, their planet-to-star flux observability peaks at long infrared wavelength, e.g. 21\,$\mu$m. We argue that the powerful exoplanet capability of JWST at those long wavelengths has been under-utilized for temperate planets. We thereby envision a third option that acquires and interprets long wavelength infrared photometry of an exoplanetary system in the ratio of two infrared wavelengths.  We call this type of observation a color phase curve (CPC).

\item \S\ref{sec: cpc} discusses that CPCs apply to non-transiting planets as well as to transiting planets, and the most natural application is to planets whose rotation is synchronous with their orbit.  For temperate rocky planets, that will often imply that the host star is an M-dwarf.  The observations we describe do not have to be continuous, rather they can be scheduled at intervals to sample peaks and valleys in the CPC, and we point out that radial velocity ephemerides are often sufficiently precise to schedule that sampling accurately.  The version of CPCs that we envision uses near-continuum bands (JWST/MIRI 12.8- and 21\,$\mu$m) to avoid possible degeneracies between spectral features and CPC amplitude.  This formulation of a CPC will detect and characterize atmospheres via their heat transfer properties, not by detecting spectral features.

\item  We consider possible sources of interference and systematic error in \S\ref{sec: interference}.  In \S\ref{sec: star}, we point out that temperature variations due to spots and faculae on the host star will cancel in the ratio of two wavelengths in the Rayleigh-Jeans limit.  Although real observations do not attain the Rayleigh-Jeans limit, the effect nevertheless greatly reduces stellar contamination due to spots and faculae.  Moreover, stellar hosts often rotate at distinctly different time scales that the planetary orbit, further facilitating separation of stellar fluctuations from the CPC measurements.  We also discuss the effect of stellar flares (\S\ref{sec: flares}), and the effect of possible changes in detector sensitivity in \S\ref{sec: instrument}. We emphasize the potential for using a single detector, as in JWST/MIRI photometry, to cancel red noise from the detector in the ratio of two wavelengths.

\item In \S\ref{sec: proxima} we illustrate how CPCs will work using the example of the Proxima Centauri system.  We simulated JWST/MIRI photometry for this system, using several models for the possible atmosphere of the 'b' planet.  We demonstrate that the atmosphere, or lack thereof, would be detectable using a well designed JWST observational campaign in the ratio of 21-to 12.8\,$\mu$m MIRI photometry.  Simulating the data, we find that CPC amplitudes of both the 'b' and 'd' planets are detectable and can be separated from low-amplitude stellar rotational variations.  

\item  We show in \S\ref{sec: contrast} that the atmospheric heat redistribution of a non-transiting planet can be determined without knowing the orbital inclination. That is possible by adopting a mass-radius relation, and by using constraints from self-consistent physical models. The detectability of an atmosphere on Proxima b using a CPC is not affected strongly by plausible variations in the heat redistribution on either the 'b' or 'd' planet (\S\ref{sec: meridional}). 

\item  Beyond temperate rocky exoplanets, CPCs have application to a wide range of both transiting and non-transiting planets, and we discuss the relevant parameter space in \S\ref{sec: sample}.  We point out that the radial velocity groups are detecting numerous new exoplanets down to the range of temperate rocky Earth-mass worlds, and most of them do not transit. New methods such as CPCs are needed to probe the atmospheres of those worlds.

\end{itemize}

\acknowledgements  We thank an anonymous referee and also Dr. Kevin Stevenson for comments that enabled us to significantly improve this paper.


\clearpage

\bibliographystyle{aasjournal}
\bibliography{references.bib}{}

@ARTICLE{krick_2016,
       author = {{Krick}, J.~E. and {Ingalls}, J. and {Carey}, S. and others},
        title = "{Spitzer IRAC Sparsely Sampled Phase Curve of the Exoplanet Wasp-14B}",
      journal = {Ap.J.},
         year = 2016,
        month = jun,
       volume = {824},
       number = {1},
          eid = {27},
        pages = {27},
          doi = {10.3847/0004-637X/824/1/27},
archivePrefix = {arXiv},
       eprint = {1603.03383},
 primaryClass = {astro-ph.EP},
       adsurl = {https://ui.adsabs.harvard.edu/abs/2016ApJ...824...27K}
}

@ARTICLE{schwartz_2015,
       author = {{Schwartz}, J.~C. and {Cowan}, N.~B.},
        title = "{Balancing the energy budget of short-period giant planets: evidence for reflective clouds and optical absorbers}",
      journal = {MNRAS},
         year = 2015,
        month = jun,
       volume = {449},
       number = {4},
        pages = {4192-4203},
          doi = {10.1093/mnras/stv470},
archivePrefix = {arXiv},
       eprint = {1502.06970},
 primaryClass = {astro-ph.EP},
       adsurl = {https://ui.adsabs.harvard.edu/abs/2015MNRAS.449.4192S}
}

@BOOK{pierrehumbert_2010,
       author = {{Pierrehumbert}, Raymond T.},
        title = "{Principles of Planetary Climate}",
         year = 2010,
         publisher= {Cambridge University Press},
       adsurl = {https://ui.adsabs.harvard.edu/abs/2010ppc..book.....P}
}

@ARTICLE{livesey_2024,
       author = {{Livesey}, Joseph R. and {Barnes}, Rory and {Deitrick}, Russell},
        title = "{Orbital Stability and Secular Dynamics of the Proxima Centauri Planetary System}",
      journal = {Ap.J.},
         year = 2024,
        month = mar,
       volume = {964},
       number = {1},
          eid = {4},
        pages = {4},
          doi = {10.3847/1538-4357/ad1ff4},
archivePrefix = {arXiv},
       eprint = {2401.08773},
 primaryClass = {astro-ph.EP},
       adsurl = {https://ui.adsabs.harvard.edu/abs/2024ApJ...964....4L}
}

@ARTICLE{revol_2024,
       author = {{Revol}, Alexandre and {Bolmont}, {\'E}meline and {Sastre}, Mariana and others},
        title = "{Drifts of the substellar points of the TRAPPIST-1 planets}",
      journal = {Astr.Ap.},
         year = 2024,
        month = nov,
       volume = {691},
          eid = {L3},
        pages = {L3},
          doi = {10.1051/0004-6361/202451532},
archivePrefix = {arXiv},
       eprint = {2409.12065},
 primaryClass = {astro-ph.EP},
       adsurl = {https://ui.adsabs.harvard.edu/abs/2024A&A...691L...3R}
}

@ARTICLE{shakespeare_2023,
       author = {{Shakespeare}, Cody J. and {Steffen}, Jason H.},
        title = "{Day and night: habitability of tidally locked planets with sporadic rotation}",
      journal = {MNRAS},
         year = 2023,
        month = oct,
       volume = {524},
       number = {4},
        pages = {5708-5724},
          doi = {10.1093/mnras/stad2162},
archivePrefix = {arXiv},
       eprint = {2303.14546},
 primaryClass = {astro-ph.EP},
       adsurl = {https://ui.adsabs.harvard.edu/abs/2023MNRAS.524.5708S}
}

@ARTICLE{johnson_2026,
       author = {{Johnson}, Ted M. and {Mandell}, Avi M.},
        title = "{A New Strategy for Using Spectroscopic Phase Curves to Characterize Nontransiting Planets}",
      journal = {Astron.J.},
         year = 2026,
        month = sep,
       volume = {172},
       number = {3},
          eid = {160},
        pages = {160},
          doi = {10.3847/1538-3881/ae8adf},
archivePrefix = {arXiv},
       eprint = {2602.07127},
 primaryClass = {astro-ph.EP},
       adsurl = {https://ui.adsabs.harvard.edu/abs/2026AJ....172..160J}
}

@ARTICLE{hammond&komacek_2024,
       author = {{Hammond}, Tobi and {Komacek}, Thaddeus D.},
        title = "{The Coupled Impacts of Atmospheric Composition and Obliquity on the Climate Dynamics of TRAPPIST-1e}",
      journal = {Ap.J.},
         year = 2024,
        month = jun,
       volume = {968},
       number = {1},
          eid = {43},
        pages = {43},
          doi = {10.3847/1538-4357/ad4a59},
archivePrefix = {arXiv},
       eprint = {2405.06615},
 primaryClass = {astro-ph.EP},
       adsurl = {https://ui.adsabs.harvard.edu/abs/2024ApJ...968...43H}
}

@ARTICLE{cowan_2007,
       author = {{Cowan}, N.~B. and {Agol}, E. and {Charbonneau}, D.},
        title = "{Hot nights on extrasolar planets: mid-infrared phase variations of hot Jupiters}",
      journal = {MNRAS},
         year = 2007,
        month = aug,
       volume = {379},
       number = {2},
        pages = {641-646},
          doi = {10.1111/j.1365-2966.2007.11897.x},
archivePrefix = {arXiv},
       eprint = {0705.1189},
 primaryClass = {astro-ph},
       adsurl = {https://ui.adsabs.harvard.edu/abs/2007MNRAS.379..641C}
}

@ARTICLE{cowan_2015,
       author = {{Cowan}, N.~B. and {Greene}, T. and {Angerhausen}, D. and others},
        title = "{Characterizing Transiting Planet Atmospheres through 2025}",
      journal = {PASP},
         year = 2015,
        month = mar,
       volume = {127},
       number = {949},
        pages = {311},
          doi = {10.1086/680855},
archivePrefix = {arXiv},
       eprint = {1502.00004},
 primaryClass = {astro-ph.EP},
       adsurl = {https://ui.adsabs.harvard.edu/abs/2015PASP..127..311C}
}

@ARTICLE{yang_2013,
       author = {{Yang}, Jun and {Cowan}, Nicolas B. and {Abbot}, Dorian S.},
        title = "{Stabilizing Cloud Feedback Dramatically Expands the Habitable Zone of Tidally Locked Planets}",
      journal = {Ap.J.(Lett.)},
         year = 2013,
        month = jul,
       volume = {771},
       number = {2},
          eid = {L45},
        pages = {L45},
          doi = {10.1088/2041-8205/771/2/L45},
archivePrefix = {arXiv},
       eprint = {1307.0515},
 primaryClass = {astro-ph.EP},
       adsurl = {https://ui.adsabs.harvard.edu/abs/2013ApJ...771L..45Y}
}

@ARTICLE{cowan_2012,
       author = {{Cowan}, Nicolas B. and {Voigt}, Aiko and {Abbot}, Dorian S.},
        title = "{Thermal Phases of Earth-like Planets: Estimating Thermal Inertia from Eccentricity, Obliquity, and Diurnal Forcing}",
      journal = {Ap.J.},
         year = 2012,
        month = sep,
       volume = {757},
       number = {1},
          eid = {80},
        pages = {80},
          doi = {10.1088/0004-637X/757/1/80},
archivePrefix = {arXiv},
       eprint = {1205.5034},
 primaryClass = {astro-ph.EP},
       adsurl = {https://ui.adsabs.harvard.edu/abs/2012ApJ...757...80C}
}

@ARTICLE{gaidos_2004,
       author = {{Gaidos}, Eric and {Williams}, D.~M.},
        title = "{Seasonality on terrestrial extrasolar planets: inferring obliquity and surface conditions from infrared light curves}",
      journal = {New Astronomy},
         year = 2004,
        month = nov,
       volume = {10},
       number = {1},
        pages = {67-77},
          doi = {10.1016/j.newast.2004.04.009},
       adsurl = {https://ui.adsabs.harvard.edu/abs/2004NewA...10...67G}
}

@ARTICLE{richardson_2003,
       author = {{Richardson}, L. Jeremy and {Deming}, Drake and {Wiedemann}, Guenter and others},
        title = "{Infrared Observations during the Secondary Eclipse of HD 209458b. I. 3.6 Micron Occultation Spectroscopy Using the Very Large Telescope}",
      journal = {Ap.J.},
         year = 2003,
        month = feb,
       volume = {584},
       number = {2},
        pages = {1053-1062},
          doi = {10.1086/345813},
archivePrefix = {arXiv},
       eprint = {astro-ph/0210612},
 primaryClass = {astro-ph},
       adsurl = {https://ui.adsabs.harvard.edu/abs/2003ApJ...584.1053R}
}

@ARTICLE{meier_valdes_2025,
       author = {{Meier Vald{\'e}s}, E.~A. and {Demory}, B.-O. and {Diamond-Lowe}, H. and others},
        title = "{Hot Rocks Survey: II. The thermal emission of TOI-1468 b reveals a bare hot rock}",
      journal = {Astr.Ap.},
         year = 2025,
        month = jun,
       volume = {698},
          eid = {A68},
        pages = {A68},
          doi = {10.1051/0004-6361/202453449},
archivePrefix = {arXiv},
       eprint = {2503.19772},
 primaryClass = {astro-ph.EP},
       adsurl = {https://ui.adsabs.harvard.edu/abs/2025A&A...698A..68M}
}

@ARTICLE{xue_2025,
       author = {{Xue}, Qiao and {Zhang}, Michael and {Coy}, Brandon Park and others},
        title = "{The JWST Rocky Worlds DDT Program Reveals GJ 3929b to Likely Be a Bare Rock}",
      journal = {Ap.J.(Lett.)},
         year = 2025,
        month = dec,
       volume = {995},
       number = {2},
          eid = {L52},
        pages = {L52},
          doi = {10.3847/2041-8213/ae2098},
archivePrefix = {arXiv},
       eprint = {2508.12516},
 primaryClass = {astro-ph.EP},
       adsurl = {https://ui.adsabs.harvard.edu/abs/2025ApJ...995L..52X}
}

@ARTICLE{august_2025,
       author = {{August}, P.~C. and {Buchhave}, L.~A. and {Diamond-Lowe}, H. and others},
        title = "{Hot Rocks Survey I: A possible shallow eclipse for LHS 1478 b}",
      journal = {Astr.Ap.},
         year = 2025,
        month = mar,
       volume = {695},
          eid = {A171},
        pages = {A171},
          doi = {10.1051/0004-6361/202452611},
archivePrefix = {arXiv},
       eprint = {2410.11048},
 primaryClass = {astro-ph.EP},
       adsurl = {https://ui.adsabs.harvard.edu/abs/2025A&A...695A.171A}
}

@ARTICLE{holmberg_2026,
       author = {{Holmberg}, M{\r{a}}ns and {Diamond-Lowe}, Hannah and {Mendon{\c{c}}a}, Jo{\~a}o M. and others},
        title = "{Hot Rocks Survey. V. Secondary Eclipse Photometry of GJ 3473 b with JWST/MIRI}",
      journal = {Astron.J.},
         year = 2026,
        month = apr,
       volume = {171},
       number = {4},
          eid = {251},
        pages = {251},
          doi = {10.3847/1538-3881/ae4c45},
archivePrefix = {arXiv},
       eprint = {2604.02332},
 primaryClass = {astro-ph.EP},
       adsurl = {https://ui.adsabs.harvard.edu/abs/2026AJ....171..251H}
}

@ARTICLE{zieba_2026,
       author = {{Zieba}, Sebastian and {Kreidberg}, Laura and {Coy}, Brandon P. and others},
        title = "{The dark and featureless surface of rocky exoplanet LHS 3844 b from JWST mid-infrared spectroscopy}",
      journal = {Nature Astronomy},
         year = 2026,
        month = may,
          doi = {10.1038/s41550-026-02860-3},
archivePrefix = {arXiv},
       eprint = {2605.00100},
 primaryClass = {astro-ph.EP},
       adsurl = {https://ui.adsabs.harvard.edu/abs/2026NatAs.tmp...97Z}
}

@ARTICLE{meadows_1996,
       author = {{Meadows}, V.~S. and {Crisp}, D.},
        title = "{Ground-based near-infrared observations of the Venus nightside: The thermal structure and water abundance near the surface}",
      journal = {JGR},
         year = 1996,
        month = jan,
       volume = {101},
       number = {E2},
        pages = {4595-4622},
          doi = {10.1029/95JE03567},
       adsurl = {https://ui.adsabs.harvard.edu/abs/1996JGR...101.4595M}
}

@ARTICLE{limbach_2024,
       author = {{Limbach}, Mary Anne and {Vanderburg}, Andrew and {Venner}, Alexander and {Blouin}, Simon and {Stevenson}, Kevin B. and {MacDonald}, Ryan J. and {Jenkins}, Sydney and {Bowens-Rubin}, Rachel and {Soares-Furtado}, Melinda and {Morley}, Caroline and {Janson}, Markus and {Debes}, John and {Xu}, Siyi and {Kleisioti}, Evangelia and {Kenworthy}, Matthew and {Butler}, Paul and {Crane}, Jeffrey D. and {Osip}, Dave and {Shectman}, Stephen and {Teske}, Johanna},
        title = "{The MIRI Exoplanets Orbiting White dwarfs (MEOW) Survey: Mid-infrared Excess Reveals a Giant Planet Candidate around a Nearby White Dwarf}",
      journal = {Ap.J.(Lett.)},
         year = 2024,
        month = sep,
       volume = {973},
       number = {1},
          eid = {L11},
        pages = {L11},
          doi = {10.3847/2041-8213/ad74ed},
archivePrefix = {arXiv},
       eprint = {2408.16813},
 primaryClass = {astro-ph.EP},
       adsurl = {https://ui.adsabs.harvard.edu/abs/2024ApJ...973L..11L}
}

@ARTICLE{lustig-yaeger_2025,
       author = {{Lustig-Yaeger}, Jacob and {Sotzen}, Kristin S. and {Stevenson}, Kevin B. and {Tsai}, Shang-Min and {Challener}, Ryan C. and {Goyal}, Jayesh and {Lewis}, Nikole K. and {Louie}, Dana R. and {Mayorga}, L.~C. and {Valentine}, Daniel and {Wakeford}, Hannah R. and {Alderson}, Lili and {Allen}, Natalie H. and {Fauchez}, Thomas J. and {Glidden}, Ana and {Gressier}, Am{\'e}lie and {H{\"o}rst}, Sarah M. and {Huang}, Jingcheng and {Lin}, Zifan and {Mandell}, Avi M. and {Mullens}, Elijah and {Peacock}, Sarah and {Schwieterman}, Edward W. and {Valenti}, Jeff A. and {Mountain}, C. Matt and {Perrin}, Marshall and {van der Marel}, Roeland P.},
        title = "{JWST-TST DREAMS: The Nightside Emission and Chemistry of WASP-17b}",
      journal = {Ap.J.(Lett.)},
         year = 2025,
        month = nov,
       volume = {994},
       number = {1},
          eid = {L4},
        pages = {L4},
          doi = {10.3847/2041-8213/ae17ae},
archivePrefix = {arXiv},
       eprint = {2510.06169},
 primaryClass = {astro-ph.EP},
       adsurl = {https://ui.adsabs.harvard.edu/abs/2025ApJ...994L...4L}
}

@INCOLLECTION{espinoza&perrin_2025,
       author = {{Espinoza}, N{\'e}stor and {Perrin}, Marshall D.},
        title = "{Highlights from Exoplanet Observations by the James Webb Space Telescope}",
    booktitle = {Handbook of Exoplanets},
         year = 2026,
    publisher = {Springer},
    editor   = {Deeg, Hans J. and Belmonte, Juan Antonio},
          eid = {216},
        pages = {216},
          doi = {10.1007/978-3-319-30648-3_216-1},
       adsurl = {https://ui.adsabs.harvard.edu/abs/2026haex.book..216E}
}

@ARTICLE{swain_2025,
       author = {{Swain}, Mark R. and {Pearson}, Kyle A. and {Komacek}, Thaddeus D. and {Bryden}, Geoffrey and {Fromont}, Emeline and {Vasisht}, Gautam and {Roudier}, Gael and {Zellem}, Robert T.},
        title = "{Thermal Phase Curves in Hot Gas Giant Exoplanets Exhibit a Complex Dependence on Planetary Properties}",
      journal = {Ap.J.},
         year = 2025,
        month = apr,
       volume = {982},
       number = {2},
          eid = {159},
        pages = {159},
          doi = {10.3847/1538-4357/adb835},
archivePrefix = {arXiv},
       eprint = {2503.00208},
 primaryClass = {astro-ph.EP},
       adsurl = {https://ui.adsabs.harvard.edu/abs/2025ApJ...982..159S}
}

@ARTICLE{newton_2018,
       author = {{Newton}, Elisabeth R. and {Mondrik}, Nicholas and {Irwin}, Jonathan and {Winters}, Jennifer G. and {Charbonneau}, David},
        title = "{New Rotation Period Measurements for M Dwarfs in the Southern Hemisphere: An Abundance of Slowly Rotating, Fully Convective Stars}",
      journal = {Astron.J.},
         year = 2018,
        month = nov,
       volume = {156},
       number = {5},
          eid = {217},
        pages = {217},
          doi = {10.3847/1538-3881/aad73b},
archivePrefix = {arXiv},
       eprint = {1807.09365},
 primaryClass = {astro-ph.SR},
       adsurl = {https://ui.adsabs.harvard.edu/abs/2018AJ....156..217N}
}

@ARTICLE{popinchalk_2021,
       author = {{Popinchalk}, Mark and {Faherty}, Jacqueline K. and {Kiman}, Rocio and {Gagn{\'e}}, Jonathan and {Curtis}, Jason L. and {Angus}, Ruth and {Cruz}, Kelle L. and {Rice}, Emily L.},
        title = "{Evaluating Rotation Periods of M Dwarfs across the Ages}",
      journal = {Ap.J.},
         year = 2021,
        month = aug,
       volume = {916},
       number = {2},
          eid = {77},
        pages = {77},
          doi = {10.3847/1538-4357/ac0444},
archivePrefix = {arXiv},
       eprint = {2105.05935},
 primaryClass = {astro-ph.SR},
       adsurl = {https://ui.adsabs.harvard.edu/abs/2021ApJ...916...77P}
}

@ARTICLE{knutson_2011,
       author = {{Knutson}, Heather A. and {Madhusudhan}, Nikku and {Cowan}, Nicolas B. and {Christiansen}, Jessie L. and {Agol}, Eric and {Deming}, Drake and {D{\'e}sert}, Jean-Michel and {Charbonneau}, David and {Henry}, Gregory W. and {Homeier}, Derek and {Langton}, Jonathan and {Laughlin}, Gregory and {Seager}, Sara},
        title = "{A Spitzer Transmission Spectrum for the Exoplanet GJ 436b, Evidence for Stellar Variability, and Constraints on Dayside Flux Variations}",
      journal = {Ap.J.},
         year = 2011,
        month = jul,
       volume = {735},
       number = {1},
          eid = {27},
        pages = {27},
          doi = {10.1088/0004-637X/735/1/27},
archivePrefix = {arXiv},
       eprint = {1104.2901},
 primaryClass = {astro-ph.EP},
       adsurl = {https://ui.adsabs.harvard.edu/abs/2011ApJ...735...27K}
}

@ARTICLE{deming_2006,
       author = {{Deming}, Drake and {Harrington}, Joseph and {Seager}, Sara and {Richardson}, L. Jeremy},
        title = "{Strong Infrared Emission from the Extrasolar Planet HD 189733b}",
      journal = {Ap.J.},
         year = 2006,
        month = jun,
       volume = {644},
       number = {1},
        pages = {560-564},
          doi = {10.1086/503358},
archivePrefix = {arXiv},
       eprint = {astro-ph/0602443},
 primaryClass = {astro-ph},
       adsurl = {https://ui.adsabs.harvard.edu/abs/2006ApJ...644..560D}
}

@ARTICLE{kaufmann_2004,
       author = {{Kaufmann}, Pierre and {Raulin}, Jean-Pierre and {de Castro}, C.~G. Gim{\'e}nez and {Levato}, Hugo and {Gary}, Dale E. and {Costa}, Joaquim E.~R. and {Marun}, Adolfo and {Pereyra}, Pablo and {Silva}, Adriana V.~R. and {Correia}, Emilia},
        title = "{A New Solar Burst Spectral Component Emitting Only in the Terahertz Range}",
      journal = {Ap.J.(Lett.)},
         year = 2004,
        month = mar,
       volume = {603},
       number = {2},
        pages = {L121-L124},
          doi = {10.1086/383186},
       adsurl = {https://ui.adsabs.harvard.edu/abs/2004ApJ...603L.121K}
}

@ARTICLE{yang_2025,
       author = {{Yang}, Xu and {Cao}, Wenda and {Wang}, Meiqi and {Jennings}, Don and {Qiu}, Jiong and {He}, Wen and {Perriyil}, Solomon M. and {Yurchyshyn}, Vasyl and {Fletcher}, Lyndsay and {Sim{\~o}es}, Paulo J.~A. and {Jhabvala}, Murzy and {Lunsford}, Allen and {Chen}, Xingyao and {Hudson}, Hugh},
        title = "{High-resolution Observations of an X6.4 Solar Flare in the Mid-infrared}",
      journal = {Ap.J.(Lett.)},
         year = 2025,
        month = aug,
       volume = {988},
       number = {2},
          eid = {L56},
        pages = {L56},
          doi = {10.3847/2041-8213/adee95},
       adsurl = {https://ui.adsabs.harvard.edu/abs/2025ApJ...988L..56Y}
}

@ARTICLE{zechmeister_2019,
       author = {{Zechmeister}, M. and {Dreizler}, S. and {Ribas}, I. and others},
        title = "{The CARMENES search for exoplanets around M dwarfs. Two temperate Earth-mass planet candidates around Teegarden's Star}",
      journal = {\aap},
         year = 2019,
        month = jul,
       volume = {627},
          eid = {A49},
        pages = {A49},
          doi = {10.1051/0004-6361/201935460},
archivePrefix = {arXiv},
       eprint = {1906.07196},
 primaryClass = {astro-ph.EP},
       adsurl = {https://ui.adsabs.harvard.edu/abs/2019A&A...627A..49Z}
}

@ARTICLE{astudillo-defru_2017,
       author = {{Astudillo-Defru}, N. and {D{\'\i}az}, R.~F. and {Bonfils}, X. and {Almenara}, J.~M. and {Delisle}, J.-B. and {Bouchy}, F. and {Delfosse}, X. and {Forveille}, T. and {Lovis}, C. and {Mayor}, M. and {Murgas}, F. and {Pepe}, F. and {Santos}, N.~C. and {S{\'e}gransan}, D. and {Udry}, S. and {W{\"u}nsche}, A.},
        title = "{The HARPS search for southern extra-solar planets. XLII. A system of Earth-mass planets around the nearby M dwarf YZ Ceti}",
      journal = {Astr.Ap.},
         year = 2017,
        month = sep,
       volume = {605},
          eid = {L11},
        pages = {L11},
          doi = {10.1051/0004-6361/201731581},
archivePrefix = {arXiv},
       eprint = {1708.03336},
 primaryClass = {astro-ph.EP},
       adsurl = {https://ui.adsabs.harvard.edu/abs/2017A&A...605L..11A}
}

@ARTICLE{bonfils_2018,
       author = {{Bonfils}, X. and {Astudillo-Defru}, N. and {D{\'\i}az}, R. and {Almenara}, J.-M. and {Forveille}, T. and {Bouchy}, F. and {Delfosse}, X. and {Lovis}, C. and {Mayor}, M. and {Murgas}, F. and {Pepe}, F. and {Santos}, N.~C. and {S{\'e}gransan}, D. and {Udry}, S. and {W{\"u}nsche}, A.},
        title = "{A temperate exo-Earth around a quiet M dwarf at 3.4 parsec}",
      journal = {Astr.Ap.},
         year = 2018,
        month = may,
       volume = {613},
          eid = {A25},
        pages = {A25},
          doi = {10.1051/0004-6361/201731973},
archivePrefix = {arXiv},
       eprint = {1711.06177},
 primaryClass = {astro-ph.EP},
       adsurl = {https://ui.adsabs.harvard.edu/abs/2018A&A...613A..25B}
}

@ARTICLE{diaz_2019,
       author = {{D{\'\i}az}, R.~F. and {Delfosse}, X. and {Hobson}, M.~J. and {Boisse}, I. and {Astudillo-Defru}, N. and {Bonfils}, X. and {Henry}, G.~W. and {Arnold}, L. and {Bouchy}, F. and {Bourrier}, V. and {Brugger}, B. and {Dalal}, S. and {Deleuil}, M. and {Demangeon}, O. and {Dolon}, F. and {Dumusque}, X. and {Forveille}, T. and {Hara}, N. and {H{\'e}brard}, G. and {Kiefer}, F. and {Lopez}, T. and {Mignon}, L. and {Moreau}, F. and {Mousis}, O. and {Moutou}, C. and {Pepe}, F. and {Perruchot}, S. and {Richaud}, Y. and {Santerne}, A. and {Santos}, N.~C. and {Sottile}, R. and {Stalport}, M. and {S{\'e}gransan}, D. and {Udry}, S. and {Unger}, N. and {Wilson}, P.~A.},
        title = "{The SOPHIE search for northern extrasolar planets. XIV. A temperate (T$_{eq}$   300 K) super-earth around the nearby star Gliese 411}",
      journal = {Astr.Ap.},
         year = 2019,
        month = may,
       volume = {625},
          eid = {A17},
        pages = {A17},
          doi = {10.1051/0004-6361/201935019},
archivePrefix = {arXiv},
       eprint = {1902.06004},
 primaryClass = {astro-ph.EP},
       adsurl = {https://ui.adsabs.harvard.edu/abs/2019A&A...625A..17D}
}

@ARTICLE{basant_2025,
       author = {{Basant}, Ritvik and {Luque}, Rafael and {Bean}, Jacob L. and {Seifahrt}, Andreas and {Brady}, Madison and {Zhao}, Lily L. and {Brown}, Nina and {Das}, Tanya and {St{\"u}rmer}, Julian and {Kasper}, David and {Gupta}, Rohan and {Stef{\'a}nsson}, Gu{\dj}mundur},
        title = "{Four Sub-Earth Planets Orbiting Barnard's Star from MAROON-X and ESPRESSO}",
      journal = {Ap.J.(Lett.)},
         year = 2025,
        month = mar,
       volume = {982},
       number = {1},
          eid = {L1},
        pages = {L1},
          doi = {10.3847/2041-8213/adb8d5},
archivePrefix = {arXiv},
       eprint = {2503.08095},
 primaryClass = {astro-ph.EP},
       adsurl = {https://ui.adsabs.harvard.edu/abs/2025ApJ...982L...1B}
}

@ARTICLE{garcia-sage_2017,
       author = {{Garcia-Sage}, K. and {Glocer}, A. and {Drake}, J.~J. and {Gronoff}, G. and {Cohen}, O.},
        title = "{On the Magnetic Protection of the Atmosphere of Proxima Centauri b}",
      journal = {Ap.J.(Lett.)},
         year = 2017,
        month = jul,
       volume = {844},
       number = {1},
          eid = {L13},
        pages = {L13},
          doi = {10.3847/2041-8213/aa7eca},
       adsurl = {https://ui.adsabs.harvard.edu/abs/2017ApJ...844L..13G}
}

@ARTICLE{collier_cameron_1999,
       author = {{Collier Cameron}, Andrew and {Horne}, Keith and {Penny}, Alan and {James}, David},
        title = "{Probable detection of starlight reflected from the giant planet orbiting {\ensuremath{\tau}} Bo{\"o}tis}",
      journal = {Nature},
         year = 1999,
        month = dec,
       volume = {402},
       number = {6763},
        pages = {751-755},
          doi = {10.1038/45451},
archivePrefix = {arXiv},
       eprint = {astro-ph/9911314},
 primaryClass = {astro-ph},
       adsurl = {https://ui.adsabs.harvard.edu/abs/1999Natur.402..751C}
}

@ARTICLE{charbonneau_1999,
       author = {{Charbonneau}, David and {Noyes}, Robert W. and {Korzennik}, Sylvain G. and {Nisenson}, Peter and {Jha}, Saurabh and {Vogt}, Steven S. and {Kibrick}, Robert I.},
        title = "{An Upper Limit on the Reflected Light from the Planet Orbiting the Star {\ensuremath{\tau}} Bootis}",
      journal = {Ap.J.(Lett.)},
         year = 1999,
        month = sep,
       volume = {522},
       number = {2},
        pages = {L145-L148},
          doi = {10.1086/312234},
archivePrefix = {arXiv},
       eprint = {astro-ph/9907195},
 primaryClass = {astro-ph},
       adsurl = {https://ui.adsabs.harvard.edu/abs/1999ApJ...522L.145C}
}

@ARTICLE{otegi_2020,
       author = {{Otegi}, J.~F. and {Bouchy}, F. and {Helled}, R.},
        title = "{Revisited mass-radius relations for exoplanets below 120 M$_{{\ensuremath{\oplus}}}$}",
      journal = {Astr.Ap.},
         year = 2020,
        month = feb,
       volume = {634},
          eid = {A43},
        pages = {A43},
          doi = {10.1051/0004-6361/201936482},
archivePrefix = {arXiv},
       eprint = {1911.04745},
 primaryClass = {astro-ph.EP},
       adsurl = {https://ui.adsabs.harvard.edu/abs/2020A&A...634A..43O}
}

@ARTICLE{klein_2021,
       author = {{Klein}, Baptiste and {Donati}, Jean-Fran{\c{c}}ois and {H{\'e}brard}, {\'E}lodie M. and {Zaire}, Bonnie and {Folsom}, Colin P. and {Morin}, Julien and {Delfosse}, Xavier and {Bonfils}, Xavier},
        title = "{The large-scale magnetic field of Proxima Centauri near activity maximum}",
      journal = {MNRAS},
         year = 2021,
        month = jan,
       volume = {500},
       number = {2},
        pages = {1844-1850},
          doi = {10.1093/mnras/staa3396},
archivePrefix = {arXiv},
       eprint = {2010.14311},
 primaryClass = {astro-ph.SR},
       adsurl = {https://ui.adsabs.harvard.edu/abs/2021MNRAS.500.1844K}
}

@ARTICLE{meadows_2018,
       author = {{Meadows}, Victoria S. and {Arney}, Giada N. and {Schwieterman}, Edward W. and {Lustig-Yaeger}, Jacob and {Lincowski}, Andrew P. and {Robinson}, Tyler and {Domagal-Goldman}, Shawn D. and {Deitrick}, Russell and {Barnes}, Rory K. and {Fleming}, David P. and {Luger}, Rodrigo and {Driscoll}, Peter E. and {Quinn}, Thomas R. and {Crisp}, David},
        title = "{The Habitability of Proxima Centauri b: Environmental States and Observational Discriminants}",
      journal = {Astrobiology},
         year = 2018,
        month = feb,
       volume = {18},
       number = {2},
        pages = {133-189},
          doi = {10.1089/ast.2016.1589},
archivePrefix = {arXiv},
       eprint = {1608.08620},
 primaryClass = {astro-ph.EP},
       adsurl = {https://ui.adsabs.harvard.edu/abs/2018AsBio..18..133M}
}

@ARTICLE{lincowski_2023,
       author = {{Lincowski}, Andrew P. and {Meadows}, Victoria S. and {Zieba}, Sebastian and {Kreidberg}, Laura and {Morley}, Caroline and {Gillon}, Micha{\"e}l and {Selsis}, Franck and {Agol}, Eric and {Bolmont}, Emeline and {Ducrot}, Elsa and {Hu}, Renyu and {Koll}, Daniel D.~B. and {Lyu}, Xintong and {Mandell}, Avi and {Suissa}, Gabrielle and {Tamburo}, Patrick},
        title = "{Potential Atmospheric Compositions of TRAPPIST-1 c Constrained by JWST/MIRI Observations at 15 {\ensuremath{\mu}}m}",
      journal = {Ap.J.(Lett.)},
         year = 2023,
        month = sep,
       volume = {955},
       number = {1},
          eid = {L7},
        pages = {L7},
          doi = {10.3847/2041-8213/acee02},
archivePrefix = {arXiv},
       eprint = {2308.05899},
 primaryClass = {astro-ph.EP},
       adsurl = {https://ui.adsabs.harvard.edu/abs/2023ApJ...955L...7L}
}

@ARTICLE{boyajian_2012,
       author = {{Boyajian}, Tabetha S. and {von Braun}, Kaspar and {van Belle}, Gerard and {McAlister}, Harold A. and {ten Brummelaar}, Theo A. and {Kane}, Stephen R. and {Muirhead}, Philip S. and {Jones}, Jeremy and {White}, Russel and {Schaefer}, Gail and {Ciardi}, David and {Henry}, Todd and {L{\'o}pez-Morales}, Mercedes and {Ridgway}, Stephen and {Gies}, Douglas and {Jao}, Wei-Chun and {Rojas-Ayala}, B{\'a}rbara and {Parks}, J. Robert and {Sturmann}, Laszlo and {Sturmann}, Judit and {Turner}, Nils H. and {Farrington}, Chris and {Goldfinger}, P.~J. and {Berger}, David H.},
        title = "{Stellar Diameters and Temperatures. II. Main-sequence K- and M-stars}",
      journal = {Ap.J.},
         year = 2012,
        month = oct,
       volume = {757},
       number = {2},
          eid = {112},
        pages = {112},
          doi = {10.1088/0004-637X/757/2/112},
archivePrefix = {arXiv},
       eprint = {1208.2431},
 primaryClass = {astro-ph.SR},
       adsurl = {https://ui.adsabs.harvard.edu/abs/2012ApJ...757..112B}
}

@ARTICLE{asplund_2005,
       author = {{Asplund}, Martin},
        title = "{New Light on Stellar Abundance Analyses: Departures from LTE and Homogeneity}",
      journal = {ARAA},
         year = 2005,
        month = sep,
       volume = {43},
       number = {1},
        pages = {481-530},
          doi = {10.1146/annurev.astro.42.053102.134001},
       adsurl = {https://ui.adsabs.harvard.edu/abs/2005ARA&A..43..481A}
}

@ARTICLE{carlsson_1992,
       author = {{Carlsson}, M. and {Rutten}, R.~J. and {Shchukina}, N.~G.},
        title = "{The formation of the MG I emission features near 12 microns}",
      journal = {Astr.Ap.},
         year = 1992,
        month = jan,
       volume = {253},
       number = {2},
        pages = {567-585},
       adsurl = {https://ui.adsabs.harvard.edu/abs/1992A&A...253..567C}
}

@ARTICLE{crossfield_2010,
       author = {{Crossfield}, Ian J.~M. and {Hansen}, Brad M.~S. and {Harrington}, Joseph and {Cho}, James Y.-K. and {Deming}, Drake and {Menou}, Kristen and {Seager}, Sara},
        title = "{A New 24 {\ensuremath{\mu}}m Phase Curve for {\ensuremath{\upsilon}} Andromedae b}",
      journal = {\apj},
         year = 2010,
        month = nov,
       volume = {723},
       number = {2},
        pages = {1436-1446},
          doi = {10.1088/0004-637X/723/2/1436},
archivePrefix = {arXiv},
       eprint = {1008.0393},
 primaryClass = {astro-ph.EP},
       adsurl = {https://ui.adsabs.harvard.edu/abs/2010ApJ...723.1436C}
}

@ARTICLE{galuzzo_2021,
       author = {{Galuzzo}, Daniele and {Cagnazzo}, Chiara and {Berrilli}, Francesco and {Fierli}, Federico and {Giovannelli}, Luca},
        title = "{Three-dimensional Climate Simulations for the Detectability of Proxima Centauri b}",
      journal = {Ap.J.},
         year = 2021,
        month = mar,
       volume = {909},
       number = {2},
          eid = {191},
        pages = {191},
          doi = {10.3847/1538-4357/abdeb4},
archivePrefix = {arXiv},
       eprint = {2102.03255},
 primaryClass = {astro-ph.EP},
       adsurl = {https://ui.adsabs.harvard.edu/abs/2021ApJ...909..191G}
}

@ARTICLE{savanov_2019,
       author = {{Savanov}, I.~S. and {Dmitrienko}, E.~S.},
        title = "{Starspot temperature}",
      journal = {INASAN Science Reports},
         year = 2019,
        month = oct,
       volume = {3},
        pages = {244-249},
          doi = {10.26087/INASAN.2019.3.1.038},
       adsurl = {https://ui.adsabs.harvard.edu/abs/2019INASR...3..244S}
}

@ARTICLE{hu_2024,
       author = {{Hu}, Renyu and {Bello-Arufe}, Aaron and {Zhang}, Michael and others},
        title = "{A secondary atmosphere on the rocky exoplanet 55 Cancri e}",
      journal = {Nature},
         year = 2024,
        month = jun,
       volume = {630},
       number = {8017},
        pages = {609-612},
          doi = {10.1038/s41586-024-07432-x},
archivePrefix = {arXiv},
       eprint = {2405.04744},
 primaryClass = {astro-ph.EP},
       adsurl = {https://ui.adsabs.harvard.edu/abs/2024Natur.630..609H}
}

@ARTICLE{teske_2025,
       author = {{Teske}, Johanna K. and {Wallack}, Nicole L. and {Piette}, Anjali A.~A. and others},
        title = "{A Thick Volatile Atmosphere on the Ultrahot Super-Earth TOI-561 b}",
      journal = {Ap.J.(Lett.)},
         year = 2025,
        month = dec,
       volume = {995},
       number = {2},
          eid = {L39},
        pages = {L39},
          doi = {10.3847/2041-8213/ae0a4c},
archivePrefix = {arXiv},
       eprint = {2509.17231},
 primaryClass = {astro-ph.EP},
       adsurl = {https://ui.adsabs.harvard.edu/abs/2025ApJ...995L..39T}
}

@ARTICLE{piaulet-ghorayeb_2025,
       author = {{Piaulet-Ghorayeb}, Caroline and {Benneke}, Bj{\"o}rn and {Turbet}, Martin and others},
        title = "{Strict Limits on Potential Secondary Atmospheres on the Temperate Rocky Exo-Earth TRAPPIST-1 d}",
      journal = {Ap.J.},
         year = 2025,
        month = aug,
       volume = {989},
       number = {2},
          eid = {181},
        pages = {181},
          doi = {10.3847/1538-4357/adf207},
archivePrefix = {arXiv},
       eprint = {2508.08416},
 primaryClass = {astro-ph.EP},
       adsurl = {https://ui.adsabs.harvard.edu/abs/2025ApJ...989..181P}
}

@ARTICLE{wargelin_2024,
       author = {{Wargelin}, Bradford J. and {Saar}, Steven H. and {Irving}, Zackery A. and others},
        title = "{X-Ray, UV, and Optical Observations of Proxima Centauri's Stellar Cycle}",
      journal = {Ap.J.},
         year = 2024,
        month = dec,
       volume = {977},
       number = {2},
          eid = {144},
        pages = {144},
          doi = {10.3847/1538-4357/ad8faa},
archivePrefix = {arXiv},
       eprint = {2411.04252},
 primaryClass = {astro-ph.SR},
       adsurl = {https://ui.adsabs.harvard.edu/abs/2024ApJ...977..144W}
}

@ARTICLE{artigau_2022,
       author = {{Artigau}, {\'E}tienne and {Cadieux}, Charles and {Cook}, Neil J. and others},
        title = "{Line-by-line Velocity Measurements: an Outlier-resistant Method for Precision Velocimetry}",
      journal = {Astron.J.},
         year = 2022,
        month = sep,
       volume = {164},
       number = {3},
          eid = {84},
        pages = {84},
          doi = {10.3847/1538-3881/ac7ce6},
archivePrefix = {arXiv},
       eprint = {2207.13524},
 primaryClass = {astro-ph.IM},
       adsurl = {https://ui.adsabs.harvard.edu/abs/2022AJ....164...84A}
}

@ARTICLE{suarez-mascareno_2020,
       author = {{Su{\'a}rez Mascare{\~n}o}, A. and {Faria}, J.~P. and {Figueira}, P. and others},
        title = "{Revisiting Proxima with ESPRESSO}",
      journal = {Astr.Ap.},
         year = 2020,
        month = jul,
       volume = {639},
          eid = {A77},
        pages = {A77},
          doi = {10.1051/0004-6361/202037745},
archivePrefix = {arXiv},
       eprint = {2005.12114},
 primaryClass = {astro-ph.EP},
       adsurl = {https://ui.adsabs.harvard.edu/abs/2020A&A...639A..77S}
}

@ARTICLE{damasso_2020,
       author = {{Damasso}, Mario and {Del Sordo}, Fabio and {Anglada-Escud{\'e}}, Guillem and others},
        title = "{A low-mass planet candidate orbiting Proxima Centauri at a distance of 1.5 AU}",
      journal = {Science Advances},
         year = 2020,
        month = jan,
       volume = {6},
       number = {3},
        pages = {eaax7467},
          doi = {10.1126/sciadv.aax7467},
       adsurl = {https://ui.adsabs.harvard.edu/abs/2020SciA....6.7467D}
}

@ARTICLE{suarez-mascareno_2025,
       author = {{Su{\'a}rez Mascare{\~n}o}, Alejandro and {Artigau}, {\'E}tienne and {Mignon}, Lucile and others},
        title = "{Diving into the planetary system of Proxima with NIRPS: Breaking the metre per second barrier in the infrared}",
      journal = {Astr.Ap.},
         year = 2025,
        month = aug,
       volume = {700},
          eid = {A11},
        pages = {A11},
          doi = {10.1051/0004-6361/202553728},
archivePrefix = {arXiv},
       eprint = {2507.21751},
 primaryClass = {astro-ph.EP},
       adsurl = {https://ui.adsabs.harvard.edu/abs/2025A&A...700A..11S}
}

@ARTICLE{kreidberg_2025,
       author = {Kreidberg, Laura and Stevenson, Kevin B.},
        title = "{A first look at rocky exoplanets with JWST}",
      journal = {arXiv e-prints},
         year = 2025,
        month = jul,
          eid = {arXiv:2507.00933},
        pages = {arXiv:2507.00933},
          doi = {10.48550/arXiv.2507.00933},
archivePrefix = {arXiv},
       eprint = {2507.00933},
 primaryClass = {astro-ph.EP},
       adsurl = {https://ui.adsabs.harvard.edu/abs/2025arXiv250700933K}
}

@ARTICLE{gillon_2025,
       author = {{Gillon}, Micha{\"e}l and {Ducrot}, Elsa and {Bell}, Taylor J. and others},
        title = "{No thick atmosphere around TRAPPIST-1 b and c from JWST thermal phase curves}",
      journal = {Nature Astronomy},
         year = 2026,
        month = may,
       volume = {10},
        pages = {674-694},
          doi = {10.1038/s41550-026-02806-9},
archivePrefix = {arXiv},
       eprint = {2509.02128},
 primaryClass = {astro-ph.EP},
       adsurl = {https://ui.adsabs.harvard.edu/abs/2026NatAs..10..674G}
}

@ARTICLE{espinoza_2025,
       author = {{Espinoza}, N{\'e}stor and {Allen}, Natalie H. and {Glidden}, Ana and others},
        title = "{JWST-TST DREAMS: NIRSpec/PRISM Transmission Spectroscopy of the Habitable Zone Planet TRAPPIST-1 e}",
      journal = {Ap.J.(Lett.)},
         year = 2025,
        month = sep,
       volume = {990},
       number = {2},
          eid = {L52},
        pages = {L52},
          doi = {10.3847/2041-8213/adf42e},
archivePrefix = {arXiv},
       eprint = {2509.05414},
 primaryClass = {astro-ph.EP},
       adsurl = {https://ui.adsabs.harvard.edu/abs/2025ApJ...990L..52E}
}

@ARTICLE{glidden_2025,
       author = {{Glidden}, Ana and {Ranjan}, Sukrit and {Seager}, Sara and others},
        title = "{JWST-TST DREAMS: Secondary Atmosphere Constraints for the Habitable Zone Planet TRAPPIST-1 e}",
      journal = {Ap.J.(Lett.)},
         year = 2025,
        month = sep,
       volume = {990},
       number = {2},
          eid = {L53},
        pages = {L53},
          doi = {10.3847/2041-8213/adf62e},
archivePrefix = {arXiv},
       eprint = {2509.05407},
 primaryClass = {astro-ph.EP},
       adsurl = {https://ui.adsabs.harvard.edu/abs/2025ApJ...990L..53G}
}

@ARTICLE{ducrot_2025,
       author = {{Ducrot}, Elsa and {Lagage}, Pierre-Olivier and {Min}, Michiel and others},
        title = "{Combined analysis of the 12.8 and 15 {\ensuremath{\mu}}m JWST/MIRI eclipse observations of TRAPPIST-1 b}",
      journal = {Nature Astronomy},
         year = 2025,
        month = mar,
       volume = {9},
        pages = {358-369},
          doi = {10.1038/s41550-024-02428-z},
archivePrefix = {arXiv},
       eprint = {2412.11627},
 primaryClass = {astro-ph.EP},
       adsurl = {https://ui.adsabs.harvard.edu/abs/2025NatAs...9..358D}
}

@ARTICLE{radica_2025,
       author = {{Radica}, Michael and {Piaulet-Ghorayeb}, Caroline and {Taylor}, Jake and others},
        title = "{Promise and Peril: Stellar Contamination and Strict Limits on the Atmosphere Composition of TRAPPIST-1 c from JWST NIRISS Transmission Spectra}",
      journal = {Ap.J.(Lett.)},
         year = 2025,
        month = jan,
       volume = {979},
       number = {1},
          eid = {L5},
        pages = {L5},
          doi = {10.3847/2041-8213/ada381},
archivePrefix = {arXiv},
       eprint = {2409.19333},
 primaryClass = {astro-ph.EP},
       adsurl = {https://ui.adsabs.harvard.edu/abs/2025ApJ...979L...5R}
}

@ARTICLE{connors_2025,
       author = {{Connors}, Nicholas J. and {Monaghan}, Christopher and {Benneke}, Bj{\"o}rn and {Dang}, Lisa},
        title = "{Uniform Reanalysis of JWST MIRI 15 {\ensuremath{\mu}}m Exoplanet Eclipse Observations Using Frame-normalized Principal Component Analysis}",
      journal = {Ap.J.(Lett.)},
         year = 2025,
        month = aug,
       volume = {989},
       number = {1},
          eid = {L11},
        pages = {L11},
          doi = {10.3847/2041-8213/adee0d},
archivePrefix = {arXiv},
       eprint = {2507.02052},
 primaryClass = {astro-ph.EP},
       adsurl = {https://ui.adsabs.harvard.edu/abs/2025ApJ...989L..11C}
}

@ARTICLE{bixel_2017,
       author = {{Bixel}, Alex and {Apai}, D{\'a}niel},
        title = "{Probabilistic Constraints on the Mass and Composition of Proxima b}",
      journal = {Ap.J.(Lett)},
         year = 2017,
        month = feb,
       volume = {836},
       number = {2},
          eid = {L31},
        pages = {L31},
          doi = {10.3847/2041-8213/aa5f51},
archivePrefix = {arXiv},
       eprint = {1702.02542},
 primaryClass = {astro-ph.EP},
       adsurl = {https://ui.adsabs.harvard.edu/abs/2017ApJ...836L..31B}
}

@ARTICLE{rackham_2018,
       author = {{Rackham}, Benjamin V. and {Apai}, D{\'a}niel and {Giampapa}, Mark S.},
        title = "{The Transit Light Source Effect: False Spectral Features and Incorrect Densities for M-dwarf Transiting Planets}",
      journal = {Ap.J.},
         year = 2018,
        month = feb,
       volume = {853},
       number = {2},
          eid = {122},
        pages = {122},
          doi = {10.3847/1538-4357/aaa08c},
archivePrefix = {arXiv},
       eprint = {1711.05691},
 primaryClass = {astro-ph.EP},
       adsurl = {https://ui.adsabs.harvard.edu/abs/2018ApJ...853..122R}
}

@ARTICLE{maurin_2012,
       author = {{Maurin}, A.~S. and {Selsis}, F. and {Hersant}, F. and {Belu}, A.},
        title = "{Thermal phase curves of nontransiting terrestrial exoplanets. II. Characterizing airless planets}",
      journal = {Astr.Ap.},
         year = 2012,
        month = feb,
       volume = {538},
          eid = {A95},
        pages = {A95},
          doi = {10.1051/0004-6361/201117054},
archivePrefix = {arXiv},
       eprint = {1110.3087},
 primaryClass = {astro-ph.EP},
       adsurl = {https://ui.adsabs.harvard.edu/abs/2012A&A...538A..95M}
}

@ARTICLE{selsis_2011,
       author = {{Selsis}, F. and {Wordsworth}, R.~D. and {Forget}, F.},
        title = "{Thermal phase curves of nontransiting terrestrial exoplanets. I. Characterizing atmospheres}",
      journal = {Astr.Ap.},
         year = 2011,
        month = aug,
       volume = {532},
          eid = {A1},
        pages = {A1},
          doi = {10.1051/0004-6361/201116654},
archivePrefix = {arXiv},
       eprint = {1104.4763},
 primaryClass = {astro-ph.EP},
       adsurl = {https://ui.adsabs.harvard.edu/abs/2011A&A...532A...1S}
}

@BOOK{natacademy_2018,
  author    = {{Charbonneau}, D. and {Gaudi}, S.},
  title     = "{Exoplanet Science Strategy}",
  isbn      = "{978-0-309-47941-7}",
  doi       = {10.17226/25187},
  url       = {https://nap.nationalacademies.org/catalog/25187/exoplanet-science-strategy},
  year      = "{2018}",
  publisher = "{The National Academies Press, Washington, DC}"
}

@INPROCEEDINGS{galuzzo_2020,
       author = {{Galuzzo}, D. and {Berrilli}, F. and {Giovannelli}, L.},
        title = "{Proxima Centauri b: infrared detectability in presence of stellar activity}",
    booktitle = {Journal of Physics Conference Series},
         year = 2020,
       series = {Journal of Physics Conference Series},
       volume = {1548},
        month = may,
    publisher = {IOP},
          eid = {012012},
        pages = {012012},
          doi = {10.1088/1742-6596/1548/1/012012},
       adsurl = {https://ui.adsabs.harvard.edu/abs/2020JPhCS1548a2012G}
}

@ARTICLE{gillon_2017,
       author = {{Gillon}, Micha{\"e}l and {Triaud}, Amaury H.~M.~J. and {Demory}, Brice-Olivier and others},
        title = "{Seven temperate terrestrial planets around the nearby ultracool dwarf star TRAPPIST-1}",
      journal = {Nature},
         year = 2017,
        month = feb,
       volume = {542},
       number = {7642},
        pages = {456-460},
          doi = {10.1038/nature21360},
archivePrefix = {arXiv},
       eprint = {1703.01424},
 primaryClass = {astro-ph.EP},
       adsurl = {https://ui.adsabs.harvard.edu/abs/2017Natur.542..456G}
}

@ARTICLE{muller_2024,
       author = {{M{\"u}ller}, Simon and {Baron}, Jana and {Helled}, Ravit and {Bouchy}, Fran{\c{c}}ois and {Parc}, L{\'e}na},
        title = "{The mass-radius relation of exoplanets revisited}",
      journal = {Astr.Ap.},
         year = 2024,
        month = jun,
       volume = {686},
          eid = {A296},
        pages = {A296},
          doi = {10.1051/0004-6361/202348690},
archivePrefix = {arXiv},
       eprint = {2311.12593},
 primaryClass = {astro-ph.EP},
       adsurl = {https://ui.adsabs.harvard.edu/abs/2024A&A...686A.296M}
}

@ARTICLE{zieba23,
       author = {{Zieba}, Sebastian and {Kreidberg}, Laura and {Ducrot}, Elsa and others},
        title = "{No thick carbon dioxide atmosphere on the rocky exoplanet TRAPPIST-1c}",
      journal = {Nature},
         year = 2023,
        month = aug,
       volume = {620},
       number = {7975},
        pages = {746-749},
          doi = {10.1038/s41586-023-06232-z},
archivePrefix = {arXiv},
       eprint = {2306.10150},
 primaryClass = {astro-ph.EP},
       adsurl = {https://ui.adsabs.harvard.edu/abs/2023Natur.620..746Z}
}

@ARTICLE{xue24,
       author = {{Xue}, Qiao and {Bean}, Jacob L. and {Zhang}, Michael and others},
        title = "{JWST Thermal Emission of the Terrestrial Exoplanet GJ 1132b}",
      journal = {Ap.J.(Lett)},
         year = 2024,
        month = sep,
       volume = {973},
       number = {1},
          eid = {L8},
        pages = {L8},
          doi = {10.3847/2041-8213/ad72e9},
archivePrefix = {arXiv},
       eprint = {2408.13340},
 primaryClass = {astro-ph.EP},
       adsurl = {https://ui.adsabs.harvard.edu/abs/2024ApJ...973L...8X}
}

@ARTICLE{mansfield24,
       author = {{Weiner Mansfield}, Megan and {Xue}, Qiao and {Zhang}, Michael and others},
        title = "{No Thick Atmosphere on the Terrestrial Exoplanet Gl 486b}",
      journal = {Ap.J.(Lett.)},
         year = 2024,
        month = aug,
        volume = {975},
          eid = {arXiv:2408.15123},
        pages = {L22},
          doi = {10.48550/arXiv.2408.15123},
archivePrefix = {arXiv},
       eprint = {2408.15123},
 primaryClass = {astro-ph.EP},
       adsurl = {https://ui.adsabs.harvard.edu/abs/2024arXiv240815123W}
}

@ARTICLE{Alderson2024,
       author = {{Alderson}, Lili and {Batalha}, Natasha E. and {Wakeford}, Hannah R. and others},
        title = "{JWST COMPASS: NIRSpec/G395H Transmission Observations of the Super-Earth TOI-836b}",
      journal = {Astron.J.},
         year = 2024,
        month = may,
       volume = {167},
       number = {5},
          eid = {216},
        pages = {216},
          doi = {10.3847/1538-3881/ad32c9},
archivePrefix = {arXiv},
       eprint = {2404.00093},
 primaryClass = {astro-ph.EP},
       adsurl = {https://ui.adsabs.harvard.edu/abs/2024AJ....167..216A}
}

@ARTICLE{LY2023a,
       author = {{Lustig-Yaeger}, Jacob and {Fu}, Guangwei and {May}, E.~M. and others},
        title = "{A JWST transmission spectrum of the nearby Earth-sized exoplanet LHS 475 b}",
      journal = {Nature Astronomy},
         year = 2023,
        month = nov,
       volume = {7},
        pages = {1317-1328},
          doi = {10.1038/s41550-023-02064-z},
archivePrefix = {arXiv},
       eprint = {2301.04191},
 primaryClass = {astro-ph.EP},
       adsurl = {https://ui.adsabs.harvard.edu/abs/2023NatAs...7.1317L}
}

@ARTICLE{MoranStevenson2023,
       author = {{Moran}, Sarah E. and {Stevenson}, Kevin B. and {Sing}, David K. and others},
        title = "{High Tide or Riptide on the Cosmic Shoreline? A Water-rich Atmosphere or Stellar Contamination for the Warm Super-Earth GJ 486b from JWST Observations}",
      journal = {Ap.J.(Lett)},
         year = 2023,
        month = may,
       volume = {948},
       number = {1},
          eid = {L11},
        pages = {L11},
          doi = {10.3847/2041-8213/accb9c},
archivePrefix = {arXiv},
       eprint = {2305.00868},
 primaryClass = {astro-ph.EP},
       adsurl = {https://ui.adsabs.harvard.edu/abs/2023ApJ...948L..11M}
}

@ARTICLE{MayMacDonald2023,
       author = {{May}, E.~M. and {MacDonald}, Ryan J. and {Bennett}, Katherine A. and others},
        title = "{Double Trouble: Two Transits of the Super-Earth GJ 1132 b Observed with JWST NIRSpec G395H}",
      journal = {Ap.J.(Lett)},
         year = 2023,
        month = dec,
       volume = {959},
       number = {1},
          eid = {L9},
        pages = {L9},
          doi = {10.3847/2041-8213/ad054f},
archivePrefix = {arXiv},
       eprint = {2310.10711},
 primaryClass = {astro-ph.EP},
       adsurl = {https://ui.adsabs.harvard.edu/abs/2023ApJ...959L...9M}
}

@ARTICLE{Kirk2024,
       author = {{Kirk}, James and {Stevenson}, Kevin B. and {Fu}, Guangwei and others},
        title = "{JWST/NIRCam Transmission Spectroscopy of the Nearby Sub-Earth GJ 341b}",
      journal = {Astron.J.},
         year = 2024,
        month = mar,
       volume = {167},
       number = {3},
          eid = {90},
        pages = {90},
          doi = {10.3847/1538-3881/ad19df},
archivePrefix = {arXiv},
       eprint = {2401.06043},
 primaryClass = {astro-ph.EP},
       adsurl = {https://ui.adsabs.harvard.edu/abs/2024AJ....167...90K}
}

@ARTICLE{Lim23,
       author = {{Lim}, Olivia and {Benneke}, Bj{\"o}rn and {Doyon}, Ren{\'e} and others},
        title = "{Atmospheric Reconnaissance of TRAPPIST-1b with JWST/NIRISS: Evidence for Strong Stellar Contamination in the Transmission Spectra}",
      journal = {Ap.J.(Lett)},
         year = 2023,
        month = sep,
       volume = {955},
       number = {1},
          eid = {L22},
        pages = {L22},
          doi = {10.3847/2041-8213/acf7c4},
archivePrefix = {arXiv},
       eprint = {2309.07047},
 primaryClass = {astro-ph.EP},
       adsurl = {https://ui.adsabs.harvard.edu/abs/2023ApJ...955L..22L}
}

@ARTICLE{Cadieux2024b,
       author = {{Cadieux}, Charles and {Doyon}, Ren{\'e} and {MacDonald}, Ryan J. and others},
        title = "{Transmission Spectroscopy of the Habitable Zone Exoplanet LHS 1140 b with JWST/NIRISS}",
      journal = {Ap.J.(Lett)},
         year = 2024,
        month = jul,
       volume = {970},
       number = {1},
          eid = {L2},
        pages = {L2},
          doi = {10.3847/2041-8213/ad5afa},
archivePrefix = {arXiv},
       eprint = {2406.15136},
 primaryClass = {astro-ph.EP},
       adsurl = {https://ui.adsabs.harvard.edu/abs/2024ApJ...970L...2C}
}

@ARTICLE{Gressier2024,
       author = {{Gressier}, Am{\'e}lie and {Espinoza}, N{\'e}stor and {Allen}, Natalie H. and others},
        title = "{Hints of a Sulfur-rich Atmosphere around the 1.6 R $_{{\ensuremath{\oplus}}}$ Super-Earth L98-59 d from JWST NIRspec G395H Transmission Spectroscopy}",
      journal = {Ap.J.(Lett.)},
         year = 2024,
        month = nov,
       volume = {975},
       number = {1},
          eid = {L10},
        pages = {L10},
          doi = {10.3847/2041-8213/ad73d1},
archivePrefix = {arXiv},
       eprint = {2408.15855},
 primaryClass = {astro-ph.EP},
       adsurl = {https://ui.adsabs.harvard.edu/abs/2024ApJ...975L..10G}
}

@ARTICLE{Damiano2024,
       author = {{Damiano}, Mario and {Bello-Arufe}, Aaron and {Yang}, Jeehyun and {Hu}, Renyu},
        title = "{LHS 1140 b Is a Potentially Habitable Water World}",
      journal = {Ap.J.(Lett)},
         year = 2024,
        month = jun,
       volume = {968},
       number = {2},
          eid = {L22},
        pages = {L22},
          doi = {10.3847/2041-8213/ad5204},
archivePrefix = {arXiv},
       eprint = {2403.13265},
 primaryClass = {astro-ph.EP},
       adsurl = {https://ui.adsabs.harvard.edu/abs/2024ApJ...968L..22D}
}

@ARTICLE{gordon_2025,
       author = {{Gordon}, Karl D. and {Sloan}, G.~C. and {Garcia Marin}, Macarena and others},
        title = "{The James Webb Space Telescope Absolute Flux Calibration. II. Mid-infrared Instrument Imaging and Coronagraphy}",
      journal = {Astron.J.},
         year = 2025,
        month = jan,
       volume = {169},
       number = {1},
          eid = {6},
        pages = {6},
          doi = {10.3847/1538-3881/ad8cd4},
archivePrefix = {arXiv},
       eprint = {2409.10443},
 primaryClass = {astro-ph.IM},
       adsurl = {https://ui.adsabs.harvard.edu/abs/2025AJ....169....6G}
}

@ARTICLE{knutson07,
   author = {{Knutson}, H.~A. and {Charbonneau}, D. and {Allen}, L.~E. and
	{Fortney}, J.~J. and {Agol}, E. and {Cowan}, N.~B. and {Showman}, A.~P. and
	{Cooper}, C.~S. and {Megeath}, S.~T.},
    title = "{A map of the day-night contrast of the extrasolar planet HD 189733b}",
  journal = {\nat},
archivePrefix = "arXiv",
   eprint = {0705.0993},
     year = 2007,
    month = may,
   volume = 447,
    pages = {183-186},
      doi = {10.1038/nature05782},
   adsurl = {http://adsabs.harvard.edu/abs/2007Natur.447..183K}
}

@ARTICLE{Scarsdale2024,
       author = {{Scarsdale}, Nicholas and {Wogan}, Nicholas and {Wakeford}, Hannah R. and others},
        title = "{JWST COMPASS: The 3{\textendash}5 {\ensuremath{\mu}}m Transmission Spectrum of the Super-Earth L 98-59 c}",
      journal = {Astron.J.},
         year = 2024,
        month = dec,
       volume = {168},
       number = {6},
          eid = {276},
        pages = {276},
          doi = {10.3847/1538-3881/ad73cf},
archivePrefix = {arXiv},
       eprint = {2409.07552},
 primaryClass = {astro-ph.EP},
       adsurl = {https://ui.adsabs.harvard.edu/abs/2024AJ....168..276S}
}

@ARTICLE{vida_2019,
       author = {{Vida}, Kriszti{\'a}n and {Ol{\'a}h}, Katalin and {K{\H{o}}v{\'a}ri}, Zsolt and {van Driel-Gesztelyi}, Lidia and {Mo{\'o}r}, Attila and {P{\'a}l}, Andr{\'a}s},
        title = "{Flaring Activity of Proxima Centauri from TESS Observations: Quasiperiodic Oscillations during Flare Decay and Inferences on the Habitability of Proxima b}",
      journal = {Ap.J.},
         year = 2019,
        month = oct,
       volume = {884},
       number = {2},
          eid = {160},
        pages = {160},
          doi = {10.3847/1538-4357/ab41f5},
archivePrefix = {arXiv},
       eprint = {1907.12580},
 primaryClass = {astro-ph.SR},
       adsurl = {https://ui.adsabs.harvard.edu/abs/2019ApJ...884..160V}
}

@ARTICLE{seager_2009,
       author = {{Seager}, S. and {Deming}, D.},
        title = "{On the Method to Infer an Atmosphere on a Tidally Locked Super Earth Exoplanet and Upper Limits to GJ 876d}",
      journal = {Ap.J.},
         year = 2009,
        month = oct,
       volume = {703},
       number = {2},
        pages = {1884-1889},
          doi = {10.1088/0004-637X/703/2/1884},
archivePrefix = {arXiv},
       eprint = {0910.1505},
 primaryClass = {astro-ph.EP},
       adsurl = {https://ui.adsabs.harvard.edu/abs/2009ApJ...703.1884S}
}

@ARTICLE{kreidberg_2016,
       author = {{Kreidberg}, Laura and {Loeb}, Abraham},
        title = "{Prospects for Characterizing the Atmosphere of Proxima Centauri b}",
      journal = {Ap.J.(Lett)},
         year = 2016,
        month = nov,
       volume = {832},
       number = {1},
          eid = {L12},
        pages = {L12},
          doi = {10.3847/2041-8205/832/1/L12},
archivePrefix = {arXiv},
       eprint = {1608.07345},
 primaryClass = {astro-ph.EP},
       adsurl = {https://ui.adsabs.harvard.edu/abs/2016ApJ...832L..12K}
}

@ARTICLE{stevenson_2020,
       author = {{Stevenson}, Kevin B. and others},
        title = "{A New Method for Studying Exoplanet Atmospheres Using Planetary Infrared Excess}",
      journal = {Ap.J.(Lett)},
         year = 2020,
        month = aug,
       volume = {898},
       number = {2},
          eid = {L35},
        pages = {L35},
          doi = {10.3847/2041-8213/aba68c},
archivePrefix = {arXiv},
       eprint = {2007.11438},
 primaryClass = {astro-ph.EP},
       adsurl = {https://ui.adsabs.harvard.edu/abs/2020ApJ...898L..35S}
}

@ARTICLE{seager_2024,
       author = {{Seager}, Sara and {Shapiro}, Alexander I.},
        title = "{Why Observations at Mid-infrared Wavelengths Partially Mitigate M Dwarf Star Host Stellar Activity Contamination in Exoplanet Transmission Spectroscopy}",
      journal = {Ap.J.},
         year = 2024,
        month = aug,
       volume = {970},
       number = {2},
          eid = {155},
        pages = {155},
          doi = {10.3847/1538-4357/ad509a},
       adsurl = {https://ui.adsabs.harvard.edu/abs/2024ApJ...970..155S}
}

@ARTICLE{ribas_2016,
       author = {{Ribas}, Ignasi and {Bolmont}, Emeline and {Selsis}, Franck and {Reiners}, Ansgar and {Leconte}, J{\'e}r{\'e}my and {Raymond}, Sean N. and {Engle}, Scott G. and {Guinan}, Edward F. and {Morin}, Julien and {Turbet}, Martin and {Forget}, Fran{\c{c}}ois and {Anglada-Escud{\'e}}, Guillem},
        title = "{The habitability of Proxima Centauri b. I. Irradiation, rotation and volatile inventory from formation to the present}",
      journal = {Astr.Ap.},
         year = 2016,
        month = dec,
       volume = {596},
          eid = {A111},
        pages = {A111},
          doi = {10.1051/0004-6361/201629576},
archivePrefix = {arXiv},
       eprint = {1608.06813},
 primaryClass = {astro-ph.EP},
       adsurl = {https://ui.adsabs.harvard.edu/abs/2016A&A...596A.111R}
}

@ARTICLE{mann_2015,
       author = {{Mann}, Andrew W. and {Feiden}, Gregory A. and {Gaidos}, Eric and {Boyajian}, Tabetha and {von Braun}, Kaspar},
        title = "{How to Constrain Your M Dwarf: Measuring Effective Temperature, Bolometric Luminosity, Mass, and Radius}",
      journal = {Ap.J.},
         year = 2015,
        month = may,
       volume = {804},
       number = {1},
          eid = {64},
        pages = {64},
          doi = {10.1088/0004-637X/804/1/64},
archivePrefix = {arXiv},
       eprint = {1501.01635},
 primaryClass = {astro-ph.SR},
       adsurl = {https://ui.adsabs.harvard.edu/abs/2015ApJ...804...64M}
}

@ARTICLE{jenkins_2019,
       author = {{Jenkins}, James S. and {Harrington}, Joseph and {Challener}, Ryan C. and others},
        title = "{Proxima Centauri b is not a transiting exoplanet}",
      journal = {MNRAS},
         year = 2019,
        month = jul,
       volume = {487},
       number = {1},
        pages = {268-274},
          doi = {10.1093/mnras/stz1268},
archivePrefix = {arXiv},
       eprint = {1905.01336},
 primaryClass = {astro-ph.EP},
       adsurl = {https://ui.adsabs.harvard.edu/abs/2019MNRAS.487..268J}
}

@ARTICLE{gilbert_2021,
       author = {{Gilbert}, Emily A. and {Barclay}, Thomas and {Kruse}, Ethan and {Quintana}, Elisa V. and {Walkowicz}, Lucianne M.},
        title = "{No Transits of Proxima Centauri planets in high cadence TESS data}",
      journal = {Frontiers in Astronomy and Space Sciences},
         year = 2021,
        month = nov,
       volume = {8},
          eid = {190},
        pages = {190},
          doi = {10.3389/fspas.2021.769371},
archivePrefix = {arXiv},
       eprint = {2110.10702},
 primaryClass = {astro-ph.EP},
       adsurl = {https://ui.adsabs.harvard.edu/abs/2021FrASS...8..190G}
}

@ARTICLE{anglada_2016,
       author = {{Anglada-Escud{\'e}}, Guillem and {Amado}, Pedro J. and {Barnes}, John and others},
        title = "{A terrestrial planet candidate in a temperate orbit around Proxima Centauri}",
      journal = {Nature},
         year = 2016,
        month = aug,
       volume = {536},
       number = {7617},
        pages = {437-440},
          doi = {10.1038/nature19106},
archivePrefix = {arXiv},
       eprint = {1609.03449},
 primaryClass = {astro-ph.EP},
       adsurl = {https://ui.adsabs.harvard.edu/abs/2016Natur.536..437A}
}

@ARTICLE{greene_2023,
       author = {{Greene}, Thomas P. and {Bell}, Taylor J. and {Ducrot}, Elsa and {Dyrek}, Achr{\`e}ne and {Lagage}, Pierre-Olivier and {Fortney}, Jonathan J.},
        title = "{Thermal emission from the Earth-sized exoplanet TRAPPIST-1 b using JWST}",
      journal = {Nature},
         year = 2023,
        month = jun,
       volume = {618},
       number = {7963},
        pages = {39-42},
          doi = {10.1038/s41586-023-05951-7},
archivePrefix = {arXiv},
       eprint = {2303.14849},
 primaryClass = {astro-ph.EP},
       adsurl = {https://ui.adsabs.harvard.edu/abs/2023Natur.618...39G}
}

\end{document}